\documentclass[aps]{revtex4}
\usepackage{amssymb}
\usepackage{amsmath}
\usepackage{graphicx}
\usepackage{latexsym}
\usepackage{bm}

\newtheorem{theorem}{Theorem}[section]

\begin{document}

\author{Sharif D. Kunikeev and Kwang S. Kim}
\affiliation{Department of Chemistry \\
Pohang University of Science and Technology \\
Pohang, 790-784, S. Korea}
\title{ {\Large Monte Carlo estimates of thermal averages and analytic
continuation}}
\date{\today}

\begin{abstract}
The Monte Carlo (MC) estimates of thermal averages are usually functions of
system control parameters $\lambda $, such as temperature, volume,
interaction couplings, etc. Given the MC average at a set of prescribed
control parameters $\lambda _{0}$, the problem of analytic continuation of
the MC data to $\lambda $-values in the neighborhood of $\lambda _{0}$ is
considered in both classic and quantum domains. The key result is the
theorem that links the differential properties of thermal averages to the
higher-order cumulants. The theorem and analytic continuation formulas
expressed via higher-order cumulants are numerically tested on the classical
Lennard-Jones cluster system of $N=13$, 55, and 147 neon particles.
\end{abstract}

\maketitle


\section{Introduction}

To obtain Monte Carlo (MC) estimates of thermal averages, one often has to
run MC codes multiple times in order to get the corresponding results at
different values of system parameters, such as temperature, volume, magnetic
or electric fields, etc., or at different values of interparticle
interaction constants. To avoid these extra time-consuming runs, several
highly effective strategies for sampling phase space of the system and/or
extracting maximum relevant information from the MC data acquired have been
suggested. Among them, these are histogram methods by Ferrenberg and
Swendsen \cite{FS:88,FS:89}, the multicanonical method by Berg and Neuhaus 
\cite{BN:91,BN:92}, the Wang-Landau method \cite{WL:01} and others.

Thus, in the first method we record a histogram $n(E)$ of how many times
each particular value of the energy $E$ is generated in MC simulation at a
particular temperature $T_{0}$. Then, using this energy distribution
histogram one can in principle recalculate the corresponding energy
distribution and thermal averages at an arbitrary temperature $T$. The
multicanonical method is based on the idea of using inverse of the density
of states, $1/\rho (E)$, instead of the Boltzmann factor $\exp (-\beta E)$,
where $\beta =1/(k_{B}T)$ and $k_{B}$ is the Boltzmann constant, for
sampling the energy states in the metastable-unstable region of the
canonical ensemble. This results in a histogram in which all energies are
sampled equally. The obvious problem with the direct implementation of this
idea is that we do not know the density of states $\rho (E)$. However, using
a sequence of approximations, Berg and Neuhaus were able to demonstrate that 
$\rho (E)$ can be found, even in a particularly impressive case of a
first-order phase transition \cite{BN:92}. On the other hand, the
Wang-Landau algorithm allows to estimate the density of states $\rho (E)$
directly instead of trying to estimate it from the probability distribution
obtained at $T_{0}$ and then with density of states one can easily calculate
the partition function, free energy, etc. at an arbitrary temperature $T$.
Moreover, the original Wang-Landau algorithm \cite{WL:01} proposed for MC
sampling in spin lattice systems has been generalized to the off-lattice
systems, such as continuum (fluid) models \cite{SDP:02,YP:03}, polymer films 
\cite{JP:02}. By a suitable reformulation of the problem the Wang-Landau
sampling scheme can be advantageously employed in quantum systems as well 
\cite{TWA:03}.

However, it should be noticed that the applicability of these methods is
severely restricted by the presence of statistical errors in the data, $n(E)$
or $\rho (E)$, generated in MC sampling. If errors in the data become
comparable or bigger than the true values in the energy range near the
average potential energy at temperature $T$, then the above methods fail to
continue the corresponding thermal averages from $T_0$ to $T$ temperatures.

In statistical physics, quantum mechanics, quantum field theory, and many
other fields of modern theoretical physics, where some kind of averaging of
a generalized exponential function is present, we constantly face the
so-called \textit{cumulant expansions }\cite{Gol:92,Ful:95,AS:06}. Although 
\textit{cumulants} (\textit{semi-invariants}) have been known long before in
mathematical statistics and probability theory, it was Kubo who first
convincingly demonstrated in his pioneering work \cite{Ku:62} how the
concept of cumulants can be widely applied to various problems of quantum
mechanics and statistical physics. For example, it has been successfully
applied to Ursell-Mayer expansion for classical and quantum gases \cite%
{MM:40} that is usually obtained by much longer diagram considerations, to
perturbation series in quantum mechanics and random perturbations in
dynamical systems, to relaxation functions in irreversible processes \cite%
{Ku:57,Ku:62}, etc. In spite of such a diversity of applications discovered
so far, cumulants and their properties have not been, to the best of our
knowledge, exploited before in the context of analytic continuation problems
both in classical and quantum domains. In his original paper, Kubo greatly
generalized the classical concept of cumulant expansion to the case of
non-commuting operator generating algebra and applied relations found in
this algebra to a diverse set of physical problems. In \cite{KF:98}, a
number of useful algebraic and geometric properties of cumulant expansions
have been summarized and applied to generate cumulant Faddeev-like equations
and to establish a method of increments for excited states. Very recently, cumulant 
expansion techniques have been successfully applied to the Fourier path integrals \cite{KFD:10,KFD:09,PKF:12}
However,
application of cumulants to the problem of analytic continuation requires
the knowledge of differential properties of cumulants. These properties have
not been established before. Therefore, one of the goals of the present
paper is to derive those properties of cumulants that are of paramount
importance for analytic continuation applications.

In this work, we present a new, more robust cumulant expansion method which
allows to analytically continue thermal averages in the neighborhood of $%
T_{0}$. In particular, we prove the key Theorem \ref{T1} which relates the
derivative of the $k$th order cumulant with respect to the inverse
temperature $\beta $ to a $(k+1)$th order cumulant. Based on this theorem,
one can develop an asymptotic expansion in the neighborhood of $T_{0}$ in
terms of higher order cumulants. Also, some numerical results for the heat
capacity of the Lennard-Jones (LJ) clusters illustrating the cumulant's
derivative theorem and the quality of the derived expansions are presented.
One of the advantages of the proposed method is that the analytic
continuation can be implemented directly without need of separate
calculating $n(E)$ or $\rho (E)$ distributions. Moreover, given the
cumulants, if necessary one can easily express these distributions in terms
of cumulants.

The rest of the paper is organized as follows. The energy representation for
the phase-space probability density function (pdf) is introduced in Section \ref{S2}. In the energy
representation, all degrees of freedom irrelevant to thermodynamic
equilibrium are integrated out. Next Section considers how partition
function, entropy, and statistical temperature can be expressed via kinetic
and potential energy pdfs. The thermodynamic energy, heat capacity and its
differential properties are given in terms of cumulant expansions in Section %
\ref{S4}. Here, the key Theorem \ref{T1} about cumulant's derivative is
formulated. In Section \ref{S5}, the cumulant expansions and the
corresponding analytic continuation formula for the energy pdf are analyzed.
The statistical or microcanonical temperature in terms of cumulants is
analyzed in Section \ref{STCE}. Further generalizations to the
multi-parameter classic and quantum systems are developed in Sections \ref%
{S6} and \ref{QM}. Here, Theorem \ref{T2}, a multi-parameter generalization
of the Theorem \ref{T1} is formulated. Numerical results are discussed in
Section \ref{RD}. Concluding remarks are in Section \ref{CR}. Finally,
technical details about cumulants, proofs of Theorems \ref{T1} and \ref{T2}
can be found in Appendices \ref{App1}-\ref{App3}.

\section{Phase space to energy space mapping}

\label{S2}

Let us consider a classical system, the phase space of which is described by
a set $\Omega$ of $3N$ canonically conjugate coordinate $\mathbf{r}=(\mathbf{%
r}_{1},\ldots ,\mathbf{r}_{N})$ and$\ 3N$ momentum $\mathbf{p}=(\mathbf{p}%
_{1},\ldots ,\mathbf{p}_{N})$ variables. On the phase space we define the
pdf $p(\Omega)$ that fully describes a thermal equilibrium state such that
the thermal average of an observable $\mathcal{O}(\Omega)$ can be calculated
as 
\begin{equation}
\langle \mathcal{O}(\Omega)\rangle \equiv \int d^{6N}\Omega\,\mathcal{O}%
(\Omega)p(\Omega)  \label{eq1}
\end{equation}%
where $d^{6N}\Omega=d^{3N}\mathbf{r}d^{3N}\mathbf{p}/w_{N}$ and $w_N$ is an
appropriate weight factor, sometimes called Gibbs factor, which makes the
classical averaging as close as possible to the quantum-mechanical one.
Further, we assume that the pdf depends on $K$ functions $H(\Omega)\equiv
(H_{1}(\Omega),\ldots ,H_{K}(\Omega))$ and $K$ control parameters $%
\lambda\equiv(\lambda _{1},\ldots ,\lambda _{K})$ so that its $\Omega$%
-dependence can be represented as %
$p(\Omega)=p(H(\Omega),\lambda)$ 
For the observable we assume a similar structure, %
$\mathcal{O}(\Omega)=\mathcal{O}(H(\Omega)),$ 
to be valid. If we define the $K$-dimensional energy pdf as%
\begin{equation}
p(E,\lambda ) = \int d^{6N}\Omega \prod\limits_{k=1}^K\delta
(E_{k}-H_{k}(\Omega)) p(H(\Omega),\lambda )  \label{eq4}
\end{equation}%
where $E\equiv(E_1,\ldots,E_K)$, then, the thermal averages can be be
evaluated by integration in the energy space only%
\begin{equation}
\langle \mathcal{O}(E)\rangle =\int d^K E\,\mathcal{O}(E)p(E,\lambda )
\label{eq5}
\end{equation}%
The original problem of integration in a $6N$-dimensional phase space is,
thus, reduced to a $K$-dimensional integration in the energy space. In Eq. (%
\ref{eq4}), we effectively integrated out all extra degrees of freedom not
important in the equilibrium state.

The problem that now can be formulated is how to calculate this energy pdf
at a fixed set of parameters. The energy pdf at an arbitrary set of
parameters $\lambda $ can be calculated, at least in principle, if it is
known at a fixed one, $\lambda _{0}$, as follows. Usually, the pdf in the
phase space is known up to the normalization factor or partition function.
We assume a generic form for 
\begin{equation}
p=\exp \left( -\lambda \cdot H(\Omega )\right) /\mathcal{Z}(\lambda )
\label{eq6}
\end{equation}%
where $\cdot $ denotes the scalar product and the partition function 
\begin{equation}
\mathcal{Z}(\lambda )=\int d^{6N}\Omega \,\exp \left( -\lambda \cdot
H(\Omega )\right)  \label{eq6a}
\end{equation}%
It assumes the existence of the partition function.

Using Metropolis \textit{et al.} importance sampling algorithm \cite%
{MRR:53,LB:05}, for which there is no need \textit{a priori} to know the
partition function, one can evaluate the integral over $\Omega $ by the
Markov chain MC simulation method and get a statistical estimate for the
energy pdf $p(E,\lambda _{0})$ at $\lambda _{0}$ parameters. To this end,
one can use a proper asymptotic representation for the $\delta $-functions
in the integrand of Eq. (\ref{eq4})%
\begin{equation}
\delta (E_{k}-H_{k})=\frac{1}{\pi }\lim\limits_{\tau \longrightarrow \infty }%
\frac{\sin \left( \tau (E_{k}-H_{k})\right) }{E_{k}-H_{k}}.  \label{eq6b}
\end{equation}

Having obtained the energy pdf at a fixed value $\lambda _{0}$, one can
easily recalculate the pdf at an arbitrary value $\lambda $ using the formula%
\begin{equation}
p(E,\lambda )=\dfrac{\exp \left( -(\lambda -\lambda_{0})\cdot E\right)
p(E,\lambda _{0})}{\mathcal{Z}(\lambda ,\lambda _{0})}  \label{eq8}
\end{equation}%
where the normalization factor%
\begin{eqnarray}
\mathcal{Z}(\lambda ,\lambda _{0}) &=&\int d^K E\,\exp \left( -(\lambda
-\lambda _{0})\cdot E\right) p(E,\lambda _{0})  \notag \\
&=&\frac{\mathcal{Z}(\lambda )}{\mathcal{Z}(\lambda _{0})}  \label{eq9}
\end{eqnarray}%
is the ratio of partition functions calculated at $\lambda $ and $\lambda
_{0}$ parameters. Thus, the thermal averages at arbitrary values of $\lambda 
$ can be evaluated with the help of Eqs. (\ref{eq5}) and (\ref{eq8}).
Examples of such calculations in the system of particles interacting via LJ
potential will be presented in Section \ref{RD}.

\section{Classical partition function, density of states and statistical
temperature}

\label{S3}

At first sight, the best one can get from Eq. (\ref{eq9}) is the ratio of
partition functions, not an absolute value at a particular set of
parameters. However, at $\lambda =0$ the partition function takes an
especially simple value: %
$\mathcal{Z}(0)=\int d^{6N}\Omega =V_{\Omega }/w_N,$ 
where $V_{\Omega }$ is the total available volume of the phase space. In
classical, non-relativistic physics $V_{\Omega }$ is, in principle, infinite
due to possible infinite particle's momentum values. However, it is well
known \cite{LL:80} that kinetic energy contribution to the partition
function, %
$\mathcal{Z}_{\mathcal{K}}(\beta )$, 
where $\beta$ is the inverse temperature, can be evaluated exactly and
calculation of the partition function, $\mathcal{Z}_{\mathcal{K}+\mathcal{V}%
}(\lambda )\equiv \mathcal{Z}_{\mathcal{K}}(\beta )\mathcal{Z}_{\mathcal{V}%
}(\lambda ),$ is, thus, reduced to the computation of the configuration
integral, $\mathcal{Z}_{\mathcal{V}}(\lambda )$, defined as in Eq. (\ref%
{eq6a}), where $\Omega \rightarrow \mathbf{r}$ is a position in the
configuration space and $H(\Omega )$ is replaced by the potential energy $%
\mathcal{V}(\mathbf{r})$.

Therefore, the configuration integral 
\begin{equation}
\mathcal{Z}_{\mathcal{V}}(0)=V^{N},  \label{eq12}
\end{equation}%
where $V$ is the spatial volume of the system. From Eq. (\ref{eq9}), we find
that%
\begin{equation}
\mathcal{Z}_{\mathcal{V}}(\lambda )=V^{N}\frac{\int d^K E\,\exp \left(
-(\lambda -\lambda_{0})\cdot E\right) p_{\mathcal{V}}(E,\lambda _{0})}{\int
d^K E\,\exp \left( \lambda _{0}\cdot E\right) p_{\mathcal{V}}(E,\lambda _{0})%
}  \label{eq13}
\end{equation}%
Here, $p_{\mathcal{V}}(E,\lambda _{0})$ is the potential energy pdf. Notice
that at $\lambda =\lambda _{0}$, the integral in the numerator is equal to
one due to normalization of the pdf, whereas at $\lambda =0$, Eq. (\ref{eq13}%
) is reduced to (\ref{eq12}).

As an example, let us consider the case of a single parameter $\lambda
_{01}=\beta _{0}=1/(k_{B}T_{0})$, where $T_{0}$ is an absolute temperature, $%
\ \mathcal{V}_{1}(\mathbf{r})=\mathcal{V}(\mathbf{r})$ is the potential
energy. We wish to calculate the density of states%
\begin{equation}
\rho _{\mathcal{K}+\mathcal{V}}(E)\equiv \int d^{6N}\Omega \,\delta (E-%
\mathcal{K}(\mathbf{p})-\mathcal{V}(\mathbf{r}))  \label{eq14}
\end{equation}%
In contrast to the partition function, the calculation of the density of
states cannot be factorized in separate integrals over momenta and
coordinates. To overcome this difficulty, it is useful first to map the
phase space to the two-dimensional energy space as suggested by Eq. (\ref%
{eq4}), namely, to define the 2$D$ pdf in a factorized form%
\begin{eqnarray}
p_{\mathcal{K},\mathcal{V}}(E_{1},E_{2},\beta _{0}) =&p_{\mathcal{K}%
}(E_{1},\beta _{0})p_{\mathcal{V}}(E_{2},\beta _{0}),  \label{eq15} \\
p_{\mathcal{K}}(E_{1},\beta _{0})=&\int d^{3N}\mathbf{p}\,\delta (E_{1}-%
\mathcal{K}(\mathbf{p}))\dfrac{\exp (-\beta _{0}\mathcal{K}(\mathbf{p}))}{%
\mathcal{Z}_{\mathcal{K}}(\beta _{0})}  \notag \\
p_{\mathcal{V}}(E_{2},\beta _{0})=& \int d^{3N}\mathbf{r}\,\delta (E_{2}-%
\mathcal{V}(\mathbf{r}))\dfrac{\exp (-\beta _{0}\mathcal{V}(\mathbf{r}))}{%
\mathcal{Z}_{\mathcal{V}}(\beta _{0})}  \notag
\end{eqnarray}%
Similar to the kinetic energy partition function, the pdf $p_{\mathcal{K}}$
can be calculated exactly; see, e.g., \cite{Ka:00}, Ch. 3. %

With the 2$D$ energy pdf (\ref{eq15}), Eq.(\ref{eq14}) can be rewritten as a
convolution of the kinetic and potential energy pdfs%
\begin{eqnarray}
\rho _{\mathcal{K}+\mathcal{V}}(E) &=&\mathcal{Z}_{\mathcal{K}+\mathcal{V}%
}(\beta _{0})\exp (\beta _{0}E)  \notag \\
&\times &\int dE_{2}\,p_{\mathcal{K}}(E-E_{2},\beta _{0})p_{\mathcal{V}%
}(E_{2},\beta _{0})  \label{eq17}
\end{eqnarray}%
Substituting explicit expressions for $\mathcal{Z}_{\mathcal{K}+\mathcal{V}}$
and $p_{\mathcal{K}}$ into (\ref{eq17}), one obtains%
\begin{eqnarray}
&&\rho _{\mathcal{K}+\mathcal{V}}(E)=\dfrac{V^{N}(2\pi m)^{\frac{3N}{2}}}{%
\Gamma (\frac{3N}{2})}  \notag \\
&\times &\dfrac{\int\limits_{-\infty }^{E}dE_{2}\,(E-E_{2})^{\frac{3N}{2}%
-1}\exp (\beta _{0}E_{2})p_{\mathcal{V}}(E_{2},\beta _{0})}{%
\int\limits_{-\infty }^{\infty }dE_{2}\,\exp (\beta _{0}E_{2})p_{\mathcal{V}%
}(E_{2},\beta _{0})},  \label{eq18}
\end{eqnarray}%
where $m$ is particle's mass and $\Gamma (\frac{3N}{2})$ the Gamma function 
\cite{AS:72}, expressed in terms of the potential energy pdf. In the
limiting case of zero interaction, $\mathcal{V}(\mathbf{r})\equiv 0$, $p_{%
\mathcal{V}\to0}(E)\to\delta (E)$ is reduced to a $\delta $-function and the
density of states Eq. (\ref{eq18}) is defined only by the kinetic energy
contribution, $\rho _{\mathcal{K}}(E)=\theta (E)V^{N}(2\pi
m)^{3N/2}E^{3N/2-1}/\Gamma (3N/2)$, where $\theta(E)$ is the Heaviside step function. It is easy to check this result directly
from Eq. (\ref{eq14}).

Moreover, from Eq. (\ref{eq18}) for the density of states we find the
corresponding expressions for the entropy, $S_{\mathcal{K}+\mathcal{V}%
}(E)=k_{B}\ln \rho _{\mathcal{K}+\mathcal{V}}(E)$, and the statistical
temperature 
\begin{eqnarray}
T_{\mathcal{K}+\mathcal{V}}(E) &=&\left( \frac{dS_{\mathcal{K}+\mathcal{V}%
}(E)}{dE}\right) ^{-1}=\dfrac{g_{1}(E)}{k_{B}\left( \frac{3N}{2}-1\right)
g_{2}(E)},  \label{eq19} \\
g_{p}(E) &=&\int\limits_{-\infty }^{E}dE_{2}\,(E-E_{2})^{\frac{3N}{2}-p}\exp
(\beta _{0}E_{2})p_{\mathcal{V}}(E_{2},\beta _{0})  \notag \\
&=&\exp(\beta_0 E)\int\limits_0^\infty dE_1 E_1^{\frac{3N}{2}%
-p}\exp(-\beta_0 E_1) p_{\mathcal{V}}(E-E_1,\beta_0)  \label{eq19a}
\end{eqnarray}%
in terms of the potential energy pdf. If $\mathcal{V}\equiv 0$, then from
Eq. (\ref{eq19}) one gets for the kinetic energy temperature 
\begin{equation}
T_{\mathcal{K}}(E)=\dfrac{\theta (E)E}{k_{B}\left( \frac{3N}{2}-1\right) }
\label{eq20}
\end{equation}%
Here, although formally $\rho _{\mathcal{K}}=0$ and $T_{\mathcal{K}}$ does
not exist at $E<0$, we put $T_{\mathcal{K}}=0$ by a continuity at negative
energies. Then, we can measure the potential energy contribution to the
temperature as the difference $T_{\mathcal{V}}\equiv T_{\mathcal{K}+\mathcal{%
V}}-T_{\mathcal{K}}$.

\section{Thermodynamic energy, heat capacity and cumulant expansion}

\label{S4}

The thermodynamic average of the energy written via the first moment or the
first cumulant term reads as%
\begin{eqnarray}
U(\beta ) &=&\frac{3N}{2\beta }+\mu _{c1}(\beta ),  \label{eq21a} \\
\mu _{c1}(\beta ) &=&\mu _{1}(\beta )=\left\langle E\right\rangle _{p_{%
\mathcal{V}}(E,\beta )}=\left\langle V(\mathbf{r})\right\rangle _{p_{%
\mathcal{V}}(\mathbf{r},\beta )}  \label{eq21b}
\end{eqnarray}%
where the first and second terms are, respectively, due to the kinetic and
potential energy contributions. Here, $\left\langle \ldots \right\rangle $
denote averages over the potential energy pdfs either in the energy or
coordinate spaces.

Differentiating the $k$th-order moment $\mu _{k}(\beta )\equiv \left\langle
E^{k}\right\rangle _{p_{\mathcal{V}}(E,\beta )}$ with respect to $\beta $,
one obtains from the definition%
\begin{equation}
\frac{d\mu _{k}}{d\beta }=-\mu _{k+1}+\mu _{k}\mu _{1}  \label{eq22}
\end{equation}%
The following theorem establishes a similar relationship between the
derivative of the $k$th and $(k+1)$th cumulants.


\begin{theorem}
\label{T1} \textrm{(univariate):} Let $\mu _{ck}(\beta )$ be the $k$th-order
classical cumulant. Then,%
\begin{equation}
\frac{d\mu _{ck}}{d\beta }=-\mu _{c(k+1)}.  \label{eq23}
\end{equation}
\end{theorem}

The general definition of the $k$th-order cumulant is given elsewhere, see,
e.g., \cite{Ku:62}, and Appendix \ref{App1}. For proof of Theorem \ref{T1}
we refer to Appendix \ref{App2}.

Using the relationship between the derivatives $d/dT=-k_{B}\beta
^{2}d/d\beta $ and Eq. (\ref{eq23}), one can easily find derivatives of
energy (\ref{eq21a}) with respect to temperature expressed in terms of the
second and higher-order cumulants. For example, for the first two
derivatives we have 
\begin{eqnarray}
C_{V} &=&\dfrac{d\tilde{U}(T)}{dT}=-k_{B}\beta ^{2}\dfrac{U(\beta )}{d\beta }%
=\frac{3}{2}Nk_{B}+k_{B}\beta ^{2}\mu _{c2}(\beta ),  \notag \\
\dfrac{dC_{V}}{dT} &=&\dfrac{d^{2}\tilde{U}(T)}{dT^{2}}=k_{B}^{2}\beta ^{3}%
\left[ -2\mu _{c2}(\beta )+\beta \mu _{c3}(\beta )\right]  \label{eq31}
\end{eqnarray}%
From (\ref{eq31}) one immediately derives that the extrema points of the
heat capacity curve are defined by equation 
\begin{equation}
2\mu _{c2}(\beta )=\beta \mu _{c3}(\beta )  \label{eq31a}
\end{equation}%
Using Taylor's series expansion near a fixed value $\beta _{0}$ and the
equation for derivatives 
\begin{equation}
\dfrac{d^{k}\mu _{cp}}{d\beta ^{k}}=(-1)^{k}\mu _{c(p+k)},\quad
p,k=1,2,3,\ldots  \label{eq31b}
\end{equation}%
that follows from Theorem \ref{T1}, Eq. (\ref{eq31a}) can be rewritten as 
\begin{equation}
\sum_{k=0}^{\infty }\left[ \beta _{0}\mu _{c(k+3)}(\beta _{0})-(k+1)\mu
_{c(k+2)}(\beta _{0})\right] \dfrac{(-\Delta \beta )^{k}}{k!}=\mu
_{c2}(\beta _{0})  \label{peak}
\end{equation}%
where $\Delta \beta =\beta -\beta _{0}$. Solving this algebraic equation
truncated at some maximum power $k=k_{max}$ for $\Delta \beta $, one obtains
a converged root that can locate an extremum position.

Similarly, for the heat capacity one obtains the expansion 
\begin{eqnarray}
\dfrac{C_{V}}{k_{B}} &=&\frac{3}{2}N+\beta ^{2}\left[ \mu _{c2}(\beta
_{0})-\mu _{c3}(\beta _{0})\Delta\beta \right.  \notag \\
&+&\left. \frac{1}{2}\mu _{c4}(\beta _{0})\Delta\beta^{2}-\frac{1}{6}\mu
_{c5}(\beta _{0})\Delta\beta^{3}+\cdots \right]  \label{eq32}
\end{eqnarray}
Some numerical applications of these equations will be given in Section \ref%
{RD}.

\section{The energy pdf and cumulant expansion}

\label{S5}

Using the Fourier integral representation for $\delta $-function, we obtain
an integral representation for 
\begin{equation}
p_{\mathcal{V}}(E,T)=\frac{1}{2\pi }\int\limits_{-\infty }^{\infty }d\tau
\exp \left( \mathrm{i}\tau E\right) \left\langle \exp \left( -\mathrm{i}\tau
V(\mathbf{r})\right) \right\rangle _{p_{\mathcal{V}}(\mathbf{r},T)}
\label{eq33}
\end{equation}

Formally, averaging the exponential function can be rewritten in terms of
cumulant's expansion as \cite{Ku:62}%
\begin{eqnarray}
\left\langle \exp \left( -\mathrm{i}\tau V(\mathbf{r})\right) \right\rangle
_{p_{\mathcal{V}}(\mathbf{r},T)}&=&\exp \left(\Xi_c(\tau,\beta) \right),
\label{eq34} \\
\Xi_c(\tau,\beta)&=&\sum\limits_{k=1}^{\infty }\frac{(-\mathrm{i}\tau )^{k}}{%
k!}\mu _{ck}(\beta )  \notag
\end{eqnarray}%
where $\mu _{ck}$ is the $k$th cumulant. On the other hand, with the help of
Taylor's series expansion and Theorem \ref{T1}, one obtains the following
cumulant expansion for 
\begin{eqnarray}
\mu_{c1}(\beta+\mathrm{i}\tau)&=&\sum_{k=0}^{\infty} \dfrac{(\mathrm{i}%
\tau)^k}{k!}\dfrac{d^k\mu_{c1}(\beta)}{d\beta^k},  \notag \\
&=&\sum_{k=0}^{\infty} \dfrac{(-\mathrm{i}\tau)^k}{k!}\mu_{c(k+1)}(\beta)
\label{eq34a}
\end{eqnarray}
Therefore, from (\ref{eq34}) and (\ref{eq34a}) one derives alternative
integral representations for 
\begin{eqnarray}
\Xi_c(\tau,\beta)=-\mathrm{i}\int_{0}^{\tau}d\tau^{\prime}\mu_{c1}(\beta+%
\mathrm{i}\tau^{\prime})  \label{eq34b}
\end{eqnarray}
and 
\begin{eqnarray}
p_{\mathcal{V}}(E,T)=\frac{1}{2\pi }\int\limits_{-\infty }^{\infty }d\tau
\exp \left( \mathrm{i}\tau E-\mathrm{i}\int_{0}^{\tau}d\tau^{\prime}%
\mu_{c1}(\beta+\mathrm{i}\tau^{\prime})\right)  \label{eq34c}
\end{eqnarray}

\begin{figure}[tbp]
\includegraphics[clip=true,width=15cm]{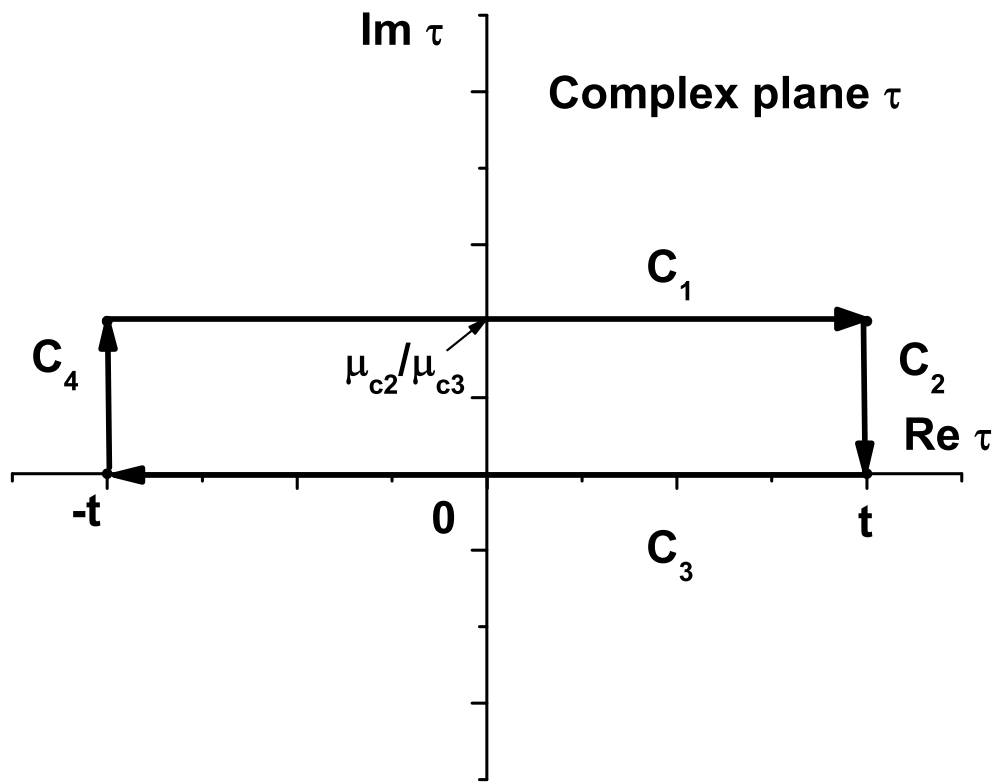}
\caption{A closed rectangular path $C=\bigcup_{i=1}^{4}C_{i}$ in the
complex plane $\tau $.}
\label{fC}
\end{figure}

Let us consider truncated cumulant expansions defined as 
\begin{equation}
\Xi _{c}^{(k_{max})}(\tau ,\beta )=\sum\limits_{k=1}^{k_{max}}\frac{(-%
\mathrm{i}\tau )^{k}}{k!}\mu _{ck}(\beta )  \label{eq34d}
\end{equation}%
for the first $k_{max}=1,2,3$ values. Thus, at $k_{max}=1$ we easily get a $%
\delta $-function like pdf 
\begin{equation}
p_{\mathcal{V}}^{(1)}=\delta (E-\mu _{c1}(\beta )),  \label{pdf1}
\end{equation}%
while at $k_{max}=2$ integration over $\tau $ yields the Gaussian function%
\begin{equation}
p_{\mathcal{V}}^{(2)}=\frac{1}{\sqrt{2\pi \mu _{c2}(\beta )}}\exp \left[ -%
\frac{(E-\mu _{c1}(\beta ))^{2}}{2\mu _{c2}(\beta )}\right] .  \label{eq35}
\end{equation}%
The next case of $k_{max}=3$ is more complicated; the pdf is reduced to an
Airy function. First, we have to shift the integration variable to the
complex plane $\tau ^{\prime }=\tau +\mathrm{i}\mu _{c2}/\mu _{c3}$ in order
to cancel the quadratic term in the exponent 
\begin{eqnarray}
p_{\mathcal{V}}^{(3)} &=&\dfrac{1}{2\pi }\int\limits_{-\infty }^{\infty
}d\tau \exp \left\{ \mathrm{i}\tau (E-\mu _{c1})-\frac{\tau ^{2}}{2}\mu
_{c2}+\mathrm{i}\frac{\tau ^{3}}{6}\mu _{c3}\right\}  \notag \\
&=&\dfrac{2^{1/3}\exp (\phi )}{2\pi \lvert \mu _{c3}\rvert ^{1/3}}%
\int\limits_{-\infty +\mathrm{i}\mu _{c2}/\mu _{c3}}^{\infty +\mathrm{i}\mu
_{c2}/\mu _{c3}}d\tau ^{\prime }\exp \left\{ \mathrm{i}\omega \tau ^{\prime
}+\mathrm{i}\frac{\tau ^{\prime }{}^{3}}{3}\right\}  \label{eq35a}
\end{eqnarray}%
where 
\begin{eqnarray}
\phi &=&\frac{\mu _{c2}}{\mu _{c3}}(E-\mu _{c1})+\frac{\mu _{c2}^{3}}{3\mu
_{c3}^{2}},  \notag \\
\omega &=&\dfrac{2^{1/3}\left( E-\mu _{c1}+\frac{\mu _{c2}^{2}}{2\mu _{c3}}%
\right) }{\mu _{c3}^{1/3}}
\end{eqnarray}%
Let us consider a closed rectangular path $C=\bigcup_{i=1}^{4}C_{i}$ in the
complex plane $\tau $ [see Fig.~\ref{fC}] consisting of two finite horizontal $C_{1}$ and $C_{3}$%
, and two vertical $C_{2}$ and $C_{4}$ segments. The horizontal segments are
defined as $C_{1}=\left\{ \tau :\,-t<\mathrm{{Re}\,\tau <t,\,{Im}\,\tau =\mu
_{c2}/\mu _{c3}}\right\} $ and $C_{3}=\left\{ \tau :\,-t<\mathrm{{Re}\,\tau
<t,\,{Im}\,\tau =0}\right\} $, where $0<t<\infty $ is a parameter fixing the
ends of segments, while the vertical segments $C_{2,4}=\left\{ \tau :\,%
\mathrm{{Re}\,\tau =\pm t,\,0<{Im}\,\tau <\mu _{c2}/\mu _{c3}}\right\} $
connect the corresponding ends of the horizontal ones. Here, we have assumed
that $C_{1}$ lies in the upper plane, i.e., $\mu _{c3}>0$; the negative case
can be considered similarly. According to the Cauchy theorem, contour
integral of a holomorphic function along a closed path is zero. Thus, we
have $\oint_{C}d\tau \cdots =\sum_{i=1}^{4}\int_{C_{i}}d\tau \cdots =0$. It
is easy to check that in the limit $t\rightarrow \infty $ contributions from
the vertical segments go to zero and, therefore, integral along the real
axis and the one taken along the path shifted into the complex plane turn
out to be the same 
\begin{equation*}
\int\limits_{-\infty +\mathrm{i}\mu _{c2}/\mu _{c3}}^{\infty +\mathrm{i}\mu
_{c2}/\mu _{c3}}d\tau \exp \left\{ \mathrm{i}\omega \tau +\mathrm{i}\frac{%
\tau ^{3}}{3}\right\} =\int\limits_{-\infty }^{\infty }d\tau \exp \left\{ 
\mathrm{i}\omega \tau +\mathrm{i}\frac{\tau ^{3}}{3}\right\}
\end{equation*}%
and (\ref{eq35a}) in terms of Airy function takes the form 
\begin{equation}
p_{\mathcal{V}}^{(3)}=\dfrac{2^{1/3}\exp (\phi )}{\lvert \mu _{c3}\rvert
^{1/3}}\,\mathrm{{Ai}\,(\omega )}  \label{Airy}
\end{equation}%
Contrary to the $p_{\mathcal{V}}^{(2)}$ pdf, which is a symmetric
distribution with respect to $\mu _{c1}$, $p_{\mathcal{V}}^{(3)}$ exhibits
an asymmetric behavior. The Airy function shows qualitatively different
behaviors: an oscillatory one at $\omega <0$, while an exponential decay at $%
\omega >0$. However, there is no guarantee that always $p_{\mathcal{V}%
}^{(3)}\geq 0$ because of possible oscillations at $\omega <0$.

One can see that the pdf $p_{\mathcal{V}}^{(k_{\max })}$ is not reducible to
elementary functions at $k_{\max }=3$. Moreover, it is expected that in
general $p_{\mathcal{V}}^{(k_{\max })}$ cannot be expressed via elementary
functions at $k_{\max }>3$ as well. At $k_{\max}\geq 3$, the asymptotic
saddle-point (SP) approximation \cite{Deb:09} can be applied in order to
derive elementary working formulas. Thus, in the SP approximation, an
integral%
\begin{equation}
I_{k_{\max }}=\mathrm{Re}\int\limits_{-\infty }^{\infty }d\tau a(\tau )\exp
\left( \Phi _{k_{\max }}(\tau )\right)  \label{Airy1}
\end{equation}%
is estimated as%
\begin{equation}
I_{k_{\max }}\sim \mathrm{Re}\sum_{\tau _{c}}a(\tau _{c})\sqrt{-\dfrac{2\pi 
}{\Phi _{k_{\max }}^{\prime \prime }(\tau _{c})}}\exp \left( \Phi _{k_{\max
}}(\tau _{c})\right)  \label{Airy2}
\end{equation}
where $\Phi _{k_{\max }}\equiv\mathrm{i}\tau E+\Xi _{c}^{(k_{max})}(\tau
,\beta )$ is a complex phase function truncated at $k_{\max } $ power and
the critical points $\tau _{c}$ are solutions of the equation $\Phi
_{k_{\max }}^{\prime }(\tau _{c})=0$.

At $k_{\max }=2$, the SP approximation reproduces the exact result (\ref%
{eq35}). Let us apply Eq. (\ref{Airy2}) to the case of $k_{\max }=3$. We
have the two critical points 
\begin{eqnarray}
\mathrm{i}\tau _{c\pm } &=&\dfrac{\mu _{c2}}{\mu _{c3}}\pm \sqrt{D},
\label{Airy3} \\
D &=&\left( \dfrac{\mu _{c2}}{\mu _{c3}}\right) ^{2}+\dfrac{2(E-\mu _{c1})}{%
\mu _{c3}}  \notag
\end{eqnarray}
The phase function $\Phi_3$ truncated at third power and its double
derivative are reduced at these points to%
\begin{eqnarray}
\Phi _3({\tau _{c\pm }}) &=&\frac{\mu _{c2}}{2}D-\frac{1}{6}\frac{\mu
_{c2}^{3}}{\mu _{c3}^{2}}\pm \frac{\mu _{c3}}{3}D^{3/2},  \label{Airy4} \\
\Phi ^{\prime \prime }_3(\tau _{c\pm }) &=&\pm \mu _{c3}\sqrt{D}  \notag
\end{eqnarray}
Substituting these equations into (\ref{Airy2}), one gets 
\begin{eqnarray}
p^{(3SP)}_\mathcal{V}&=&\mathrm{Re}\,\left\{\sqrt{\dfrac{1}{2\pi\mu_{c3}%
\sqrt{D}}} \exp\left(\dfrac{\mu_{c2}}{2}D-\dfrac{\mu_{c2}^3}{\mu_{c3}^2}%
\right)\right.  \notag \\
&\times&\left. \left[\mathrm{i}\exp\left(\dfrac{\mu_{c3}}{3}D^{3/2} \right)
+ \exp\left(-\dfrac{\mu_{c3}}{3}D^{3/2} \right)\right] \right\}
\label{Airy5}
\end{eqnarray}
The SP approximation (\ref{Airy5}) is seen to have a $1/4$th power
singularity at $D=0$ or at $E_{sing}=\mu_{c1}-\mu_{c2}^2/(2\mu_{c3})$. At
the singularity point, where the two critical points coalesce, the standard
SP approximation is not valid.

If the pdf $p_{\mathcal{V}}(\beta_0)$ is known at a fixed value $\beta_0$,
then its analytic continuation to $\beta=\beta_0+\Delta\beta$ point is given
by 
\begin{eqnarray}
p_{\mathcal{V}}(E,\beta )=\dfrac{\exp \left( -\Delta\beta E\right) p_{%
\mathcal{V}}(E,\beta _{0})} {S(\Delta\beta,\beta_0)}  \label{eq36}
\end{eqnarray}
where the normalization factor 
\begin{eqnarray}
S(\Delta\beta,\beta_0)=\int d E\,\exp \left( -\Delta\beta E\right) p_{%
\mathcal{V}}(E,\beta _{0})
\end{eqnarray}
Let us calculate the normalization for the first three truncated pdfs, $p_{%
\mathcal{V}}^{(1,2,3)}$. One gets consecutively 
\begin{eqnarray}
S^{(1)}&=&\exp\left(-(\Delta\beta)\mu_{c1}(\beta_0)\right),  \notag \\
S^{(2)}&=&\exp\left(-(\Delta\beta)\mu_{c1}(\beta_0)+\frac{\mu_{c2}(\beta_0)}{%
2}(\Delta\beta)^2\right),  \notag \\
S^{(3)}&=&\exp\left(-(\Delta\beta)\mu_{c1}(\beta_0)+\frac{\mu_{c2}(\beta_0)}{%
2}(\Delta\beta)^2-\frac{\mu_{c3}(\beta_0)}{6}(\Delta\beta)^3\right)
\label{eq37}
\end{eqnarray}
where in the last line the Airy averaging has been carried out with the help
of an integral formula \cite{VS:04} 
\begin{equation}
\int_{-\infty}^\infty dt\,\exp(pt)\mathrm{Ai}\,(t)=\exp(p^3/3), \quad 
\mathrm{Re}\,p>0.
\end{equation}
If the average potential energy $\mu_{c1}$ is negative, then the
normalization factor is exponentially small or large, depending on the sign
of $\Delta\beta$. At $\Delta\beta > 0$, $S$ is large or it is small
otherwise.

Eqs. (\ref{eq37}) suggest an anzatz for 
\begin{eqnarray}
S(\Delta\beta,\beta_0)&=&\exp\left(\kappa(\Delta\beta,\beta_0)\right), 
\notag \\
\kappa(\Delta\beta,\beta_0)&=&\sum_{k=1}^\infty c_k(\beta_0)(\Delta\beta)^k
\end{eqnarray}
where the first three coefficients are defined by Eq. (\ref{eq37}) as $%
c_k(\beta_0)=(-1)^{k+1}\mu_{ck}(\beta_0)/k!$, $k=1,2,3$. It can be easily
checked that the same formula holds valid for the higher-order coefficients
at $k>3$. It follows directly from equation 
\begin{eqnarray}
\mu_{c2}(\Delta\beta,\beta_0)&=&\dfrac{d^2
S(\Delta\beta,\beta_0)/d(\Delta\beta)^2}{S(\Delta\beta,\beta_0)}-\left(%
\dfrac{d S(\Delta\beta,\beta_0)/d(\Delta\beta)}{S(\Delta\beta,\beta_0)}%
\right)^2  \notag \\
&=&\dfrac{d^2 \kappa(\Delta\beta,\beta_0)}{d(\Delta\beta)^2}
\end{eqnarray}
as a result of equating the like powers in Taylor's expansions for $%
\mu_{c2}(\Delta\beta,\beta_0)$ [see Eq. (\ref{eq32})] and the second
derivative of $\kappa(\Delta\beta,\beta_0)$ in the right-hand side.

\section{The Statistical Temperature and cumulant expansion}

\label{STCE}

Making use of the integral representation Eq. (\ref{eq34c}), the statistical
temperature in terms of cumulants can be rewritten as 
\begin{eqnarray}
T_{\mathcal{K+V}}(E) &=&\dfrac{f_{1}(E)}{k_{B}f_{2}(E)},  \label{ST1} \\
f_{p}(E) &=&\int_{-\infty }^{\infty }d\tau a_{p}(\tau )\exp (\Phi (\tau ))
\label{ST1a}
\end{eqnarray}%
where 
\begin{eqnarray}
a_{p}(\tau ) &=&(\beta _{0}+\mathrm{i}\tau )^{-\alpha _{p}},\quad \alpha
_{p}=3N/2-p+1,  \notag \\
\Phi (\tau ) &=&\mathrm{i}\tau E-\mathrm{i}\int_{0}^{\tau }d\tau ^{\prime
}\mu _{c1}(\beta _{0}+\mathrm{i}\tau ^{\prime })  \notag \\
&=&\mathrm{i}\tau E+\sum\limits_{k=1}^{\infty }\frac{\mu _{ck}(\beta _{0})}{%
k!}(-\mathrm{i}\tau )^{k}  \label{ST1b}
\end{eqnarray}%
The second line in $\Phi (\tau )$ is due to Theorem \ref{T1}. $f_{p}(E)$ can
be written in an explicitly real form%
\begin{equation}
f_{p}(E)=2\int_{0}^{\infty }d\tau \frac{\exp \left( g(\tau )\right) }{\left(
\beta _{0}^{2}+\tau ^{2}\right) ^{\alpha _{p}/2}}\cos \left( \varphi (\tau
)\right)  \label{ST2}
\end{equation}%
where%
\begin{eqnarray}
g(\tau ) &=&\sum\limits_{s=1}^{\infty }\frac{\mu _{c(2s)}(\beta _{0})}{(2s)!}%
\left( -\tau ^{2}\right) ^{s},  \notag \\
\varphi (\tau ) &=&E\tau -\sum\limits_{s=0}^{\infty }\frac{\mu
_{c(2s+1)}(\beta _{0})}{(2s+1)!}(-1)^{s}\tau ^{2s+1}-\alpha _{p}\arctan 
\frac{\tau }{\beta _{0}}  \label{ST3}
\end{eqnarray}%
While the 'real' form (\ref{ST2}) might be convenient for numerical
estimates of the integrals, the 'complex' representation (\ref{ST1a}) is a
good starting point for developing the SP approximations.  The critical
points $\tau _{c}$ are defined by 
\begin{equation}
\sum_{k=0}^{k_{max} }\dfrac{\mu _{c(k+1)}(\beta _{0})}{k!}(-\mathrm{i}\tau
_{c})^{k}=E  \label{ST4}
\end{equation}%
Solving Eq. (\ref{ST4}) for $\tau _{c}$, one gets the critical points. In
the simplest approximation, truncating equation at $k_{max}=1$, we find 
\begin{equation}
\mathrm{i}\tau _{c}=-\dfrac{E-\mu _{c1}(\beta _{0})}{\mu _{c2}(\beta _{0})}.
\end{equation}%
Then, with the help of (\ref{Airy2}) one obtains an asymptotic estimate for
the temperature 
\begin{eqnarray}
T_{\mathcal{K+V}}(E) &=&\dfrac{T_{0}}{1-\dfrac{k_{B}T_{0}(E-\mu _{c1}(\beta
_{0}))}{\mu _{c2}(\beta _{0})}}  \notag \\
&\approx &T_{0}\left[ 1+\dfrac{k_{B}T_{0}(E-\mu _{c1}(\beta _{0}))}{\mu
_{c2}(\beta _{0})}\right]  \label{ST5}
\end{eqnarray}%
where the second line is valid if the ratio in the denominator taken by
modulus is much less than one. In the neighborhood of $\mu _{c1}$, $T_{%
\mathcal{K+V}}$ is seen to grow linearly as $E$ increases. Notice that a
similar piecewise linear interpolation scheme for the statistical
temperature has been suggested in the statistical-temperature MC (STMC) and
molecular dynamics (STMD) algorithms \cite{KSK:06}.

At $k_{max}=2$, solving the quadratic equation one obtains the two critical
points (\ref{Airy3}). Substituting the phase function and its double
derivative (\ref{ST1b}) taken at these points  into (\ref{Airy2}), where $%
a(\tau)=(\beta_0+\mathrm{i}\tau)^{-\alpha_p}$, one gets%
\begin{equation}
f_{p}(E)\sim \mathrm{Re}\,\sum_{\pm }\frac{1}{\left( \beta _{0}+\dfrac{\mu
_{c2}}{\mu _{c3}}\pm \sqrt{D}\right) ^{\alpha _{p}}}\sqrt{\frac{-1}{\pm \mu
_{c3}\sqrt{D}}}\exp \left( \pm \frac{\mu _{c3}}{3}D^{3/2}\right)  \label{ST9}
\end{equation}
Let us assume that $E$ is in the neighborhood of $\mu_{c1}$ such that $D>0$.
Then, depending on the sign of $\mu_{c3}$, we find that one term in the sum (%
\ref{ST9}) is real and the other purely imaginary. Thus, if $\mu_{c3}>0$,
the "$-$"-sign term is real, while the "$+$"-sign term is imaginary. Taking
into account only the real contributions to $f_p$, one obtains 
\begin{eqnarray}
T_{\mathcal{K+V}}(E) &=&\dfrac{1}{k_B\left(\beta_0+\dfrac{\mu_{c2}}{\mu_{c3}}
-\mathrm{sign}\,(\mu_{c3})\sqrt{D} \right)}  \label{ST10}
\end{eqnarray}
where $\mathrm{sign}\,(x)=\pm 1$ if $x$ is positive or negative. Moreover,
if 
\begin{equation*}
\left|\dfrac{2(E-\mu_{c1})}{\mu_{c3}}\right|\ll\left(\dfrac{\mu_{c2}}{%
\mu_{c3}}\right)^2
\end{equation*}
and 
\begin{equation*}
\sqrt{D}\approx \left|\dfrac{\mu_{c2}}{\mu_{c3}} \right|+\mathrm{sign}%
\,(\mu_{c3}) \dfrac{E-\mu_{c1}}{\mu_{c2}},
\end{equation*}
then (\ref{ST10}) reduces to (\ref{ST5}).

\section{Multivariate Cumulant Expansions}

\label{S6}

It is straightforward to generalize the above results to the multivariate
case. Thus, using Fourier integral representation for $\delta $-functions in
the energy pdf (\ref{eq4}), we obtain 
\begin{eqnarray}
p(E,\lambda ) &=&\dfrac{1}{(2\pi )^{K}}\idotsint\limits_{-\infty }^{\infty
}d\tau _{1}\cdots d\tau _{K}\exp \left( \mathrm{i}\sum_{k=1}^{K}E_{k}\tau
_{k}\right) \left\langle \exp \left( -\mathrm{i}\sum_{k=1}^{K}H_{k}(\Omega
)\tau _{k}\right) \right\rangle _{p(\Omega ,\lambda )}  \notag \\
&=&\dfrac{1}{(2\pi )^{K}}\idotsint\limits_{-\infty }^{\infty }d\tau
_{1}\cdots d\tau _{K}\exp \left( \mathrm{i}\sum_{k=1}^{K}E_{k}\tau
_{k}+\sum_{s_{1},\ldots ,s_{K}}\dfrac{(-\mathrm{i}\tau _{1})^{s_{1}}\cdots (-%
\mathrm{i}\tau _{K})^{s_{K}}}{s_{1}!\cdots s_{K}!}\mu _{cs_{1}\ldots
s_{K}}\right)
\end{eqnarray}%
where summation runs over non-negative integers $(s_{1},\ldots ,s_{K})\neq
(0,\ldots ,0)$. The pdf in the phase space, \ $p(\Omega ,\lambda ),$\ is
defined by Eq. (\ref{eq6}). Multivariate moments can be defined as averages
either over the energy or the phase space pdfs: $\mu _{s_{1}\ldots
s_{K}}=\left\langle E_{1}^{s_{1}}\cdots E_{K}^{s_{K}}\right\rangle
_{p(E,\lambda )}=\left\langle H_{1}^{s_{1}}\cdots H_{K}^{s_{K}}\right\rangle
_{p(\Omega ,\lambda )}$. Relationships between multivariate moments and
cumulants $\mu _{cs_{1}\ldots s_{K}}$ can be established using the
moment-generating function [see Appendix \ref{App3}]. Truncating the
cumulant expansion in the exponent by quadratic terms, $s_{1}+\ldots
+s_{K}\leq 2$, the $K$-dimensional Gaussian integral evaluates to%
\begin{equation}
p^{GK}(E,\lambda )=\frac{1}{\left[ (2\pi )^{K}\det \sigma \right] ^{1/2}}%
\exp \left[ -\frac{1}{2}\sum_{k,k^{\prime }=1}^{K}(E_{k}-\bar{E}_{k})\left(
\sigma ^{-1}\right) _{kk^{\prime }}(E_{k^{\prime }}-\bar{E}_{k^{\prime }})%
\right]
\end{equation}%
where%
\begin{eqnarray}
\bar{E}_{k}(\lambda ) &=&\mu _{\substack{ 0\ldots 010\ldots 0  \\ k}}=\mu 
_{\substack{ c0\ldots 010\ldots 0  \\ \,\,\,\,k}}=\left\langle
H_{k}\right\rangle _{p(\Omega ,\lambda )},  \notag \\
\sigma _{kk^{\prime }}(\lambda ) &=&\mu _{\substack{ c0\ldots 010\ldots
010\ldots 0  \\ \quad \,\,\,\,\,\,k\quad \,\,\,\,k^{\prime }\quad }}%
=\left\langle H_{k}H_{k^{\prime }}\right\rangle _{p(\Omega ,\lambda
)}-\left\langle H_{k}\right\rangle _{p(\Omega ,\lambda )}\left\langle
H_{k^{\prime }}\right\rangle _{p(\Omega ,\lambda )}
\end{eqnarray}

The multivariate analogue of Theorem \ref{T1} on cumulant's derivative is

\begin{theorem}
\label{T2} \textrm{(multivariate):} Let $\mu _{cs_1\ldots s_r\ldots s_K}
(\lambda_1,\ldots,\lambda_r,\ldots,\lambda_K )$ be a $K$-variate cumulant.
Then,%
\begin{equation}
\frac{\partial\mu _{c\ldots s_r\ldots}}{\partial\lambda_r }%
(\ldots\lambda_r\ldots)=-\mu _{c\ldots(s_r+1)\ldots}(\ldots\lambda_r\ldots).
\end{equation}
\end{theorem}

Proof of this equation is similar to that done in the univariate case [see
Appendix \ref{App3}].

As an example, let us consider an application of this equation to the LJ
cluster system. Let $N$ particles interacting via LJ potential $\mathcal{V}%
_{LJ}$ be put in a cubic thermostatic container of size $L$ and volume $V=L^3
$; the system is kept at temperature $T$. The free energy of system 
\begin{eqnarray}
F=F_{ideal}-k_B T\ln\dfrac{1}{V^N}\idotsint d^{3N}\mathbf{r}\,
\exp\left(-\beta \mathcal{V}_{LJ}(\mathbf{r}) \right)
\end{eqnarray}
where $F_{ideal}$ is the free-energy of an ideal, non-interacting system and 
$d^{3N}\mathbf{r}=d^3 \mathbf{r}_1\ldots d^3\mathbf{r}_N$. It is convenient
to rescale the LJ potential to the size of container $\mathbf{r}\to\mathbf{u}%
=\mathbf{r}/L$ separately for the repulsive and attractive parts: 
\begin{eqnarray}
\beta\mathcal{V}_{LJ}(\mathbf{r})&=&\beta V_1(\mathbf{r})+\beta V_2(\mathbf{r%
})  \notag \\
&=&4\beta\varepsilon_{LJ} \sum_{i<j=1}^N\left(\dfrac{\sigma_{LJ}}{\lvert%
\mathbf{r}_i-\mathbf{r}_j\rvert}\right)^{12} -4\beta\varepsilon_{LJ}
\sum_{i<j=1}^N\left(\dfrac{\sigma_{LJ}}{\lvert\mathbf{r}_i-\mathbf{r}_j\rvert%
}\right)^{6}  \notag \\
&=&\lambda_1 V_1(\mathbf{u})+\lambda_2 V_2(\mathbf{u})
\end{eqnarray}
where $\varepsilon_{LJ}$ and $\sigma_{LJ}$ are the standard LJ energy and
length parameters and 
\begin{eqnarray}
V_1(\mathbf{u})&=&\sum_{i<j=1}^N\dfrac{1}{\lvert\mathbf{u}_i-\mathbf{u}%
_j\rvert^{12}}, \quad \lambda_1=4\varepsilon_{LJ}\beta\left(\dfrac{V_{LJ}}{V}%
\right)^4,  \notag \\
V_2(\mathbf{u})&=&-\sum_{i<j=1}^N\dfrac{1}{\lvert\mathbf{u}_i-\mathbf{u}%
_j\rvert^{6}}, \quad \lambda_2=4\varepsilon_{LJ}\beta\left(\dfrac{V_{LJ}}{V}%
\right)^2.
\end{eqnarray}
Here, $V_{LJ}=\sigma_{LJ}^3$ denotes an effective LJ volume. Notice that (i)
in the rescaled variables integration over $\mathbf{u}$ in the configuration
integral is carried out in a $3N$-dimensional hypercube of unit size and
(ii) all the dependences on system parameters $T$ and $V$ are included in
the dimensionless parameters $\lambda\equiv(\lambda_1,\lambda_2)$. Taking
derivative of the free energy with respect to $V$ at fixed $T$ and $N$, we
obtain pressure 
\begin{eqnarray}
p=-\left(\dfrac{\partial F}{\partial V}\right)_{T,N}=\dfrac{Nk_B T}{V}\left[%
1+\dfrac{4\lambda_1}{N} \mu_{c10}(\lambda)+ \dfrac{2\lambda_2}{N}
\mu_{c01}(\lambda)\right]  \label{6.1}
\end{eqnarray}
expressed in terms of the first order cumulants 
\begin{eqnarray}
\mu_{c10}(\lambda)&=&\left\langle V_1\right\rangle_{p(\mathbf{u},\lambda)}= 
\dfrac{\idotsint\limits _0^1 d^{3N}\mathbf{u}\,V_1(\mathbf{u})
\exp(-\lambda_1 V_1(\mathbf{u})-\lambda_2 V_2(\mathbf{u}))}{\idotsint\limits
_0^1 d^{3N}\mathbf{u}\,\exp(-\lambda_1 V_1(\mathbf{u})-\lambda_2 V_2(\mathbf{%
u}))},  \notag \\
\mu_{c01}(\lambda)&=&\left\langle V_2\right\rangle_{p(\mathbf{u},\lambda)}= 
\dfrac{\idotsint\limits _0^1 d^{3N}\mathbf{u}\,V_2(\mathbf{u})
\exp(-\lambda_1 V_1(\mathbf{u})-\lambda_2 V_2(\mathbf{u}))}{\idotsint\limits
_0^1 d^{3N}\mathbf{u}\,\exp(-\lambda_1 V_1(\mathbf{u})-\lambda_2 V_2(\mathbf{%
u}))}  \label{6.2}
\end{eqnarray}
Note that $\mu_{c10}$ and $\mu_{c01}$ are universal functions, in the sense
that their functional dependence is universal, not depending on the specific
potential interaction parameters $\sigma_{LJ}$ and $\varepsilon_{LJ}$, as
well as the system parameters $T$ and $V$. The first term in (\ref{6.1}) is
the pressure of an ideal system with no interaction between particles. The
second (positive) and the third (negative) terms are \textit{predominantly}
contributions due to repulsive and attractive interactions, respectively.
Strictly speaking, the second (third) term contains contributions from both
repulsive and attractive interactions, but effects due to repulsive
(attractive) interactions are expected to be dominant.

Making use of Theorem \ref{T2}, one can expand cumulants in Taylor's series
around a fixed value $\lambda_0\equiv(\lambda_{10},\lambda_{20})$: 
\begin{eqnarray}
\mu_{c10}(\lambda)&=&\mu_{c10}(\lambda_0)-\mu_{c20}(\lambda_0)\Delta%
\lambda_1-\mu_{c11}(\lambda_0)\Delta\lambda_2  \notag \\
&+&\frac{1}{2}\mu_{c30}(\lambda_0)(\Delta\lambda_1)^2+
\mu_{c21}(\lambda_0)\Delta\lambda_1\Delta\lambda_2+ \frac{1}{2}%
\mu_{c12}(\lambda_0)(\Delta\lambda_2)^2+\ldots,  \notag \\
\mu_{c01}(\lambda)&=&\mu_{c01}(\lambda_0)-\mu_{c11}(\lambda_0)\Delta%
\lambda_1-\mu_{c02}(\lambda_0)\Delta\lambda_2  \notag \\
&+&\frac{1}{2}\mu_{c21}(\lambda_0)(\Delta\lambda_1)^2+
\mu_{c12}(\lambda_0)\Delta\lambda_1\Delta\lambda_2+ \frac{1}{2}%
\mu_{c03}(\lambda_0)(\Delta\lambda_2)^2+\ldots  \label{6.3}
\end{eqnarray}
where $\Delta\lambda_1=\lambda_1-\lambda_{10}$, $\Delta\lambda_2=\lambda_2-%
\lambda_{20}$ and the higher-order cumulants are explicitly defined by 
\begin{eqnarray}
\mu_{c20}&=&\left\langle V_1^2 \right\rangle -\left\langle V_1
\right\rangle^2,\quad \mu_{c11}=\left\langle V_1 V_2 \right\rangle
-\left\langle V_1 \right\rangle \left\langle V_2 \right\rangle,  \notag \\
\mu_{c02}&=&\left\langle V_2^2 \right\rangle -\left\langle V_2
\right\rangle^2,\quad \mu_{c30}=\left\langle V_1^3 \right\rangle
-3\left\langle V_1^2 \right\rangle \left\langle V_1 \right\rangle
+2\left\langle V_1 \right\rangle^3,  \notag \\
\mu_{c21}&=&\left\langle V_1^2 V_2 \right\rangle-2\left\langle V_1 V_2
\right\rangle \left\langle V_1 \right\rangle-\left\langle V_1^2
\right\rangle\left\langle V_2 \right\rangle+ 2\left\langle V_1
\right\rangle^2\left\langle V_2 \right\rangle,  \notag \\
\mu_{c12}&=&\left\langle V_1 V_2^2 \right\rangle-2\left\langle V_1 V_2
\right\rangle \left\langle V_2 \right\rangle-\left\langle V_1
\right\rangle\left\langle V_2^2 \right\rangle+ 2\left\langle V_1
\right\rangle\left\langle V_2 \right\rangle^2,  \notag \\
\mu_{c03}&=&\left\langle V_2^3 \right\rangle -3\left\langle V_2^2
\right\rangle \left\langle V_2 \right\rangle +2\left\langle V_2
\right\rangle^3, \cdots  \label{6.4}
\end{eqnarray}
For brevity, we dropped pdf's indication on the averaging operation. With
the help of Eqs. (\ref{6.1})-(\ref{6.4}), one can analytically continue the
state equation in the neighborhood of $\lambda_0$.

The critical point $(T_{c},V_{c})$ is defined by equations 
\begin{equation}
\left( \dfrac{\partial p}{\partial V}\right) _{T,N}=0,\quad \left( \dfrac{%
\partial ^{2}p}{\partial V^{2}}\right) _{T,N}=0.
\end{equation}%
Written in terms of cumulants, these equations for the LJ-system are
equivalent to 
\begin{eqnarray}
16\lambda _{1}^{2}\mu _{c20}+16\lambda _{1}\lambda _{2}\mu _{c11}+4\lambda
_{2}^{2}\mu _{c02} &=&N+20\lambda _{1}\mu _{c10}+6\lambda _{2}\mu _{c01}, 
\notag \\
64\lambda _{1}^{3}\mu _{c30}+96\lambda _{1}^{2}\lambda _{2}\mu
_{c21}+48\lambda _{1}\lambda _{2}^{2}\mu _{c12}+8\lambda _{2}^{3}\mu _{c03}
&=&208\lambda _{1}^{2}\mu _{c20}+28\lambda _{2}^{2}\mu _{c02}+64\lambda
_{1}\lambda _{2}\mu _{c11}  \notag \\
&-&80\lambda _{1}\mu _{c10}-12\lambda _{2}\mu _{c01}
\end{eqnarray}%
Again, expanding cumulants in Taylor's series around a $\lambda _{0}$-value,
which is supposed to be close to the critical value $\lambda _{c}\equiv
(\lambda _{1c},\lambda _{2c})$, one gets a system of two-variate polynomial
equations. Once a critical solution $\lambda _{c}$ of these equations is
found, the critical volume and temperature are given by 
\begin{equation}
V_{c}=V_{LJ}\sqrt{\dfrac{\lambda _{2c}}{\lambda _{1c}}},\quad T_{c}=\dfrac{%
4\varepsilon _{LJ}}{k_{B}}\dfrac{\lambda _{1c}}{\lambda _{2c}^{2}}.
\end{equation}

\section{Quantum Generalizations}

\label{QM}

In quantum mechanics, the classical pdfs are substituted by the statistical
density operators $\hat{\rho}=\exp (-\beta \hat{H})$, where $\hat{H}$ is the
system Hamiltonian operator, so that the quantum thermodynamic average is
defined by 
\begin{equation}
\langle \hat{\mathcal{O}}\hat{\rho}\rangle =\mathrm{tr}\,(\hat{\mathcal{O}}%
\hat{\rho})/\mathrm{tr}\,\hat{\rho}  \label{7.1}
\end{equation}%
where $\hat{\mathcal{O}}$ is an observable operator. Using the path-integral
representation for the density operator, a quantum system can be effectively
mapped to a corresponding classical [polymer-type] statistical system. Based
on this mapping, we can develop similar analytic continuation methods in
quantum domain as well. Thus, the usual Feynman path integral expression 
\cite{FH:65} for the density matrix reads as an integral over all curves
connecting the two configurations $\mathbf{x}$ and $\mathbf{x}^{\prime }$:

\begin{eqnarray}
\rho (\mathbf{x},\mathbf{x}^{\prime };\beta ) &\equiv &\left\langle \mathbf{x%
}^{\prime }\right\vert \exp (-\beta \hat{H})\left\vert \mathbf{x}%
\right\rangle  \notag \\
&=&\idotsint\limits_{\mathbf{x}(\tau ):\,\mathbf{x}(0)=\mathbf{x,\,x}(\beta
\hbar )=\mathbf{x}^{\prime }}\mathcal{D}\left[ \mathbf{x}(\tau )\right] \exp
\left\{ -\frac{1}{\hbar }S\left[ \mathbf{x}(\tau );\beta \right] \right\} ,
\label{q1} \\
S\left[ \mathbf{x}(\tau );\beta \right] &=&\int\limits_{0}^{\beta \hbar
}d\tau H\left[ \mathbf{x}(\tau )\right] =\int\limits_{0}^{\beta \hbar }d\tau
\left\{ \frac{1}{2}m\mathbf{\dot{x}}(\tau )^{2}+V\left[ \mathbf{x}(\tau )%
\right] \right\}  \label{q2}
\end{eqnarray}%
The symbol $\mathcal{D}\left[ \mathbf{x}(\tau )\right] $ indicates that the
integration is performed over a set of all continuous, non-differentiable
[zigzag-type] curves $\mathbf{x}(\tau ):\left[ 0,\beta \hbar \right]
\rightarrow \mathbf{R}^{d}$, with $\,\mathbf{x}(0)=\mathbf{x}$ and $\mathbf{x%
}(\beta \hbar )=\mathbf{x}^{\prime }$; $\hbar $\ is Planck's constant. The
integer $d$ reflects the dimensionality,\ with $d=3N$ for a system of $N$
particles having a mass $m$ \ and interacting via a potential $V$ in
3-dimensional space.

Calculating the path integral is a challenging task, which in general cannot
be performed analytically. It is only for simple model problems, such as
quadratic potentials, that an exact solution can be obtained. For more
complex systems, the path integral has traditionally been estimated using
the discretized time-slicing approximation \cite{Cep:95} or "Fourier
discretization" \cite{Mil:75,FD:84}. By introducing a change of variables to
simplify the boundary conditions and temperature dependence: $\mathbf{x}%
(\tau )=\mathbf{x}+(\mathbf{x}^{\prime }-\mathbf{x})\tau /(\beta \hbar )+%
\mathbf{y}(\tau /(\beta \hbar ))$, the reduced paths given by $\mathbf{y}$,
will satisfy Dirichlet boundary conditions, $\mathbf{y}(0)=\mathbf{y}(1)=0$,
independent of $\mathbf{x}$, $\mathbf{x}^{\prime }$, and $\beta $. In the
Fourier path representation, Cartesian components $y_{i}$, $i=1,\ldots ,d$
of $\mathbf{y}$ are expanded\ in a complete set of sinusoidal basis
functions 
\begin{equation}
y_{i}(u)=\sum_{k=1}^{\infty }a_{ki}\Lambda _{k}(u),\quad \Lambda _{k}(u)=%
\sqrt{2}\dfrac{\sin (k\pi u)}{k\pi }  \label{7.2}
\end{equation}%
where coefficients of the Fourier expansion, $\left\{ a_{ki}\right\} $, are
new functional integral variables. Or, in vector notations, we write $%
\mathbf{y}(u)=\sum_{k=1}^{\infty }\mathbf{a}_{k}\Lambda _{k}(u)$, where $%
\mathbf{a}_{k}=(a_{k1},\ldots ,a_{kd})$. In these variables, Eq. (\ref{q1})
can be rewritten in the form%
\begin{eqnarray}
\rho (\mathbf{x},\mathbf{x}^{\prime };\beta ) &=&\lim_{K\rightarrow \infty
}\rho ^{(K)}(\mathbf{x},\mathbf{x}^{\prime };\lambda ),  \notag \\
\rho ^{(K)}(\mathbf{x},\mathbf{x}^{\prime };\lambda ) &\equiv &\left( \dfrac{%
\lambda _{1}}{2\pi }\right) ^{(K+1)d/2}\idotsint\limits_{-\infty }^{\infty }d%
\mathbf{a}_{1}\cdots d\mathbf{a}_{K}\exp \left( -\lambda
_{1}S_{1}^{(K)}-\lambda _{2}S_{2}^{(K)}\right) ,  \label{7.3} \\
S_{1}^{(K)} &=&\frac{1}{2}\left[ (\mathbf{x}^{\prime }-\mathbf{x}%
)^{2}+\sum\limits_{k=1}^{K}\mathbf{a}_{k}^{2}\right] ,  \label{7.4} \\
S_{2}^{(K)} &=&\int\limits_{0}^{1}du\,V\left[ \mathbf{x}+(\mathbf{x}^{\prime
}-\mathbf{x})u+\sum_{k=1}^{K}\mathbf{a}_{k}\Lambda _{k}(u)\right] ,
\label{7.5}
\end{eqnarray}%
where $\lambda _{1}=1/\sigma ^{2}$, $\sigma =\left( \beta \hbar
^{2}/m\right) ^{1/2}$, and $\lambda _{2}=\beta $. The $\sigma $ parameter
differs from the usual thermal de Broglie wavelength at the corresponding
temperature by a factor of $\sqrt{2}$. $S_{1}^{(K)}$ and $S_{2}^{(K)}$ are
contributions to the action from the kinetic and potential energy operators,
respectively. Observe that both $S_{1}^{(K)}$ and $S_{2}^{(K)}$ do not
depend on $\beta $; all the dependence on $\beta $ is separated out in the $%
\lambda _{1,2}$ parameters: $\lambda _{1(2)}$ is inversely (directly)
proportional to $\beta $. Truncated at first $K$ vector functional variables 
$\mathbf{a}_{1},\ldots ,\mathbf{a}_{K}$, the $\rho ^{(K)}$ \ is said to be
the density matrix in the primitive Fourier (PF) approximation. As $%
K\rightarrow \infty $, $\rho ^{(K)}$ approaches an exact value $\rho $ at
the convergence rate $1/K$ \cite{EDC:99}.

The structure of $\rho ^{(K)}$ is seen to be very similar to that of the
classical pdfs and, therefore, we can apply the above analytic continuation
techniques for thermodynamic averages developed in the classical case. For
example, let us consider a quantum estimator for the thermodynamic energy
which can be obtained from the system partition function 
\begin{equation}
\mathcal{Z}^{(K)}(\lambda )=\int d\mathbf{x}\,\rho ^{(K)}(\mathbf{x},\mathbf{%
x};\lambda ).  \label{7.6}
\end{equation}%
The expression for the energy is given by 
\begin{eqnarray}
U(\beta ) &=&-\dfrac{\partial \ln \mathcal{Z}^{(K)}}{\partial \beta }=-%
\dfrac{\partial \lambda _{1}}{\partial \beta }\dfrac{1}{\mathcal{Z}^{(K)}}%
\dfrac{\partial \mathcal{Z}^{(K)}}{\partial \lambda _{1}}-\dfrac{\partial
\lambda _{2}}{\partial \beta }\dfrac{1}{\mathcal{Z}^{(K)}}\dfrac{\partial 
\mathcal{Z}^{(K)}}{\partial \lambda _{2}}  \notag \\
&=&\frac{(K+1)d}{2\beta }-\frac{\lambda _{1}}{\beta }\mu _{c10}(\lambda
)+\mu _{c01}(\lambda )  \label{7.7}
\end{eqnarray}%
where the first order cumulants%
\begin{eqnarray}
\mu _{c10} &=&\left\langle S_{1}^{(K)}\right\rangle _{p_{q}(\mathbf{X}%
,\lambda )},\quad \mu _{c01}=\left\langle S_{2}^{(K)}\right\rangle _{p_{q}(%
\mathbf{X},\lambda )},  \notag \\
p_{q}(\mathbf{X},\lambda ) &=&\frac{\exp \left( -\lambda _{1}S_{1}^{(K)}(%
\mathbf{X})-\lambda _{2}S_{2}^{(K)}(\mathbf{X})\right) }{\int d\mathbf{X\,}%
\exp \left( -\lambda _{1}S_{1}^{(K)}(\mathbf{X})-\lambda _{2}S_{2}^{(K)}(%
\mathbf{X})\right) }  \label{7.8}
\end{eqnarray}%
Here, $\mathbf{X}$ labels the whole set of integration variables $\mathbf{%
X\equiv }(\mathbf{x},\mathbf{a}_{1},\cdots ,\mathbf{a}_{K})$. Observe that
at $\mathbf{x}=\mathbf{x}^{\prime }$, $S_{1}^{(K)}$ does not depend on $%
\mathbf{x}$. In the case of zero potential energy $V\equiv 0$, one obtains
that $\mu _{c10}=dK/(2\lambda _{1})$, $\mu _{c01}=0$ and $U=d/(2\beta )$.

Moreover, making use of Theorem \ref{T2}, one can easily derive a quantum
estimator for the heat capacity%
\begin{equation}
\frac{C_{V}}{k_{B}}=\frac{1}{2}(K+1)d-2\lambda _{1}\mu _{c10}+\lambda
_{1}^{2}\mu _{c20}-2\lambda _{1}\beta \mu _{c11}+\beta ^{2}\mu _{c02}
\label{7.9}
\end{equation}%
In the limit of zero potential, it is easy to check that $\mu _{c11}=\mu
_{c02}=0$, $\mu _{c20}=-\partial \mu _{c10}/\partial \lambda
_{1}=dK/(2\lambda _{1}^{2})$ and we get $C_{V}/k_{B}=d/2$, the value
expected for an ideal gas. Similar to expansions (\ref{6.3}), Theorem \ref%
{T2} can further be used to develop Taylor's series expansions of Eqs. (\ref%
{7.7}) and (\ref{7.9}) around a fixed value $\lambda _{0}$ in terms of the
higher-order cumulants and, thus, to get an analytic continuation of the
thermodynamic averages in $\beta $ parameter.

Note that quantum estimators of the type (\ref{7.7}) and (\ref{7.9}) might
be more advantageous in MC simulations since they require only knowledge of
potential functions as opposed to those obtained, e.g., in \cite%
{PSD:03a,PSD:03b}, which require first and second derivatives of the
potential energy to be calculated. It is straightforward with the help of
Theorem \ref{T2} to obtain similar analytic continuation formulas in terms
of higher order cumulants in the discretized time-slicing primitive
approximation \cite{Cep:95}.

\section{Results and Discussion}

\label{RD}

To illustrate the above analytic continuation formulas we consider as
testing system a cluster of $N=13$ neon atoms interacting via LJ potential,
with the corresponding standard LJ length and energy parameters $\sigma
_{LJ}=2.749$ \textrm{{\AA } }and $\epsilon _{LJ}=35.6$ K being used. The
mass of the Ne atom was set to $m=20.0$, the rounded atomic mass of the most
abundant isotope. The particles are assumed to be into the sphere with the
confining radius\textrm{\ }$R_{c}=0.85\sigma _{LJ}N^{1/3}=2.0\sigma _{LJ}$.

\begin{figure}[tbp]
\includegraphics[clip=true,width=15cm]{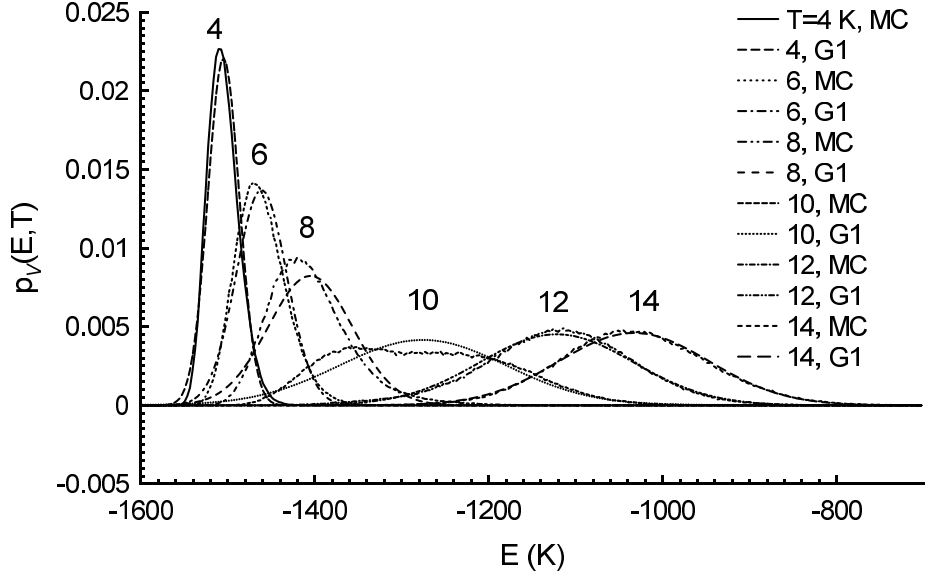}
\caption{The potential energy pdfs $p_{\mathcal{V}}(E,T)$ for the system of $%
N=13$ neon particles with atomic mass $m=20.0$ a.u. interacting via
Lennard-Jones potential with the parameters $\protect\sigma_{LJ}=2.749$ 
\textrm{{\AA } }and $\protect\epsilon _{LJ}=35.6$ K as a function of energy $%
E$ calculated at fixed temperatures $T=4$, 6, 8, 10, 12, and 14 K. The
particles are assumed to be in the sphere with the confining radius\textrm{\ 
}$R_{c}=2.0\protect\sigma _{LJ}$. MC labels the total results of MC
simulations using the asymptotic formula (\protect\ref{eq6b}) at $\protect%
\tau=10^2$ for $\protect\delta$-function, whereas G1 is the Gaussian
approximation Eq. (\protect\ref{eq35}) to the MC pdf.}
\label{f1}
\end{figure}

In Fig. \ref{f1}, the potential energy pdfs $p_{\mathcal{V}}(E,T)$ (labeled
by MC) are plotted at fixed temperatures $T=4,6,8,10,12,$ and 14 K. The MC
simulations were implemented using the parallel tempering technique, also
known as replica exchange Markov chain MC sampling \cite%
{SW:86,Gey:91,GT:95,SO:99,ED:05,KTH:06,SMF:08,BNJ:08,BJ:11}. In this method
several replicas of the same system are simulated in parallel in the
canonical ensemble, and usually each replica at a different temperature. In
this work, 29 replicas have been run on the even temperature grid, with the
temperature step $\Delta T=1$ K, in the interval from 3 to 31 K. Parallel
tempering is complementary to any set of MC moves for a system at a single
temperature, and such single-system moves are performed between each
attempted swap of complete configurations of the systems at adjacent
temperatures. The swap moves have been attempted randomly with the
probability $P_{swap}=0.1/N$. The high temperature systems are generally
able to sample larger volumes of the configuration space, whereas low
temperature systems may become trapped in local energy minima. Thus,
swapping of configurations ensures that the lower temperature systems can
access all contributing regions of the configuration integral, thereby
overcoming potential barriers between the local energy minima.

The asymptotic formula (\ref{eq6b}), with the asymptotic parameter $\tau $
set to 10$^{2}$, has been employed to calculate the $\delta $-function. The
energy spectra have been calculated on the equidistant grid of $10^{3}$
points in the range $\left[ -2000,1000\right] $ K. The total \ number of MC
moves has been divided into 50 statistical blocks; the first block data
being far from equilibration, have been discarded in further averaging. In
each block, the number of MC moves has been $N_{bl}=10^{5}N$. The pdfs in
the Gaussian approximation Eq. (\ref{eq35}), labeled by G1, are also shown
in Fig. \ref{f1}. Observe that the Gaussian approximation reproduces MC pdfs
quite well at higher temperatures $T\geq 12$ K, but there both quantitative
and qualitative differences in the shape of distributions are seen at lower
temperatures, especially at $T=8$ and 10 K, where the "melting peak" in heat
capacity $C_{V}$ is formed. It should be noticed that although $\delta $%
-function is a non-negative function (distribution), its asymptotic
approximation (\ref{eq6b}) can in principle be negative at finite values of $%
\tau $. Thus, although MC pdfs in Fig \ref{f1} are seen to be apparently
non-negative, we found that in the regions far from the central peaks, where
its values are totally corrupted by statistical errors, the MC pdf can take
very small negative values $\sim -10^{-6}-10^{-8}$. In these regions
negative MC values should be just zeroed. Also,  note that 
 usage of a Gaussian representation 
for the $\delta$-function can be a guarantee for non-negativity of the pdf 
[suggestion of an anonymous referee].

\begin{figure}[tbp]
\includegraphics[clip=true,width=15cm]{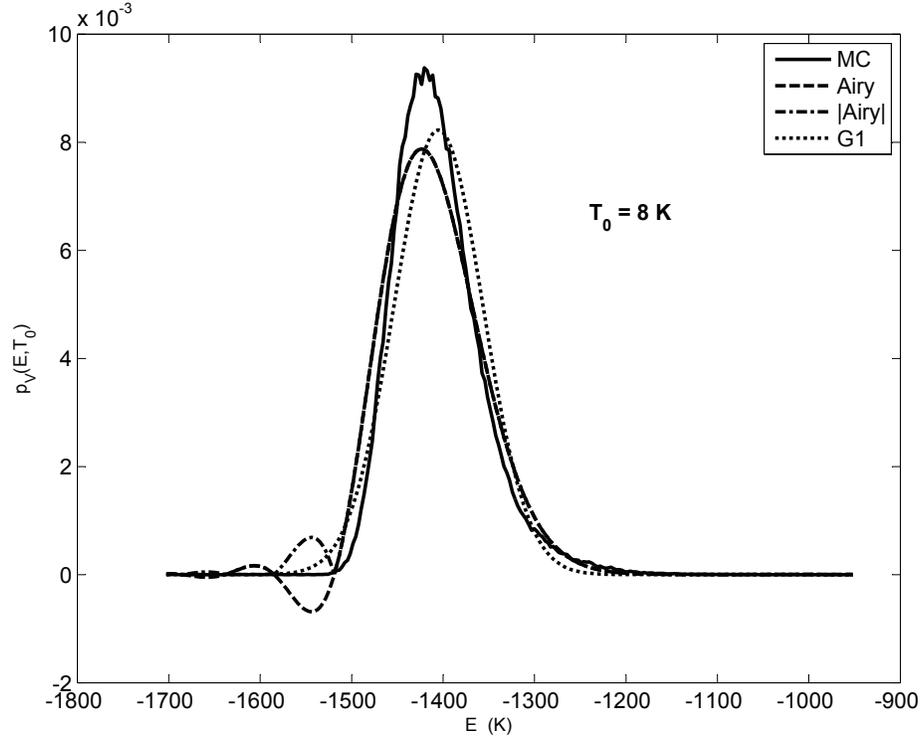}
\caption{Effect of the 3rd cumulant term on the pdf at $T_0=8 $ K. Here,
'Airy' labels the Airy pdf Eq. (\protect\ref{Airy}), whereas '$|\mathrm{Airy}%
|$' is the modulus of the Airy pdf. The system and other notations are the
same as in Fig. \protect\ref{f1}.}
\label{f2}
\end{figure}

In Fig. \ref{f2}, the results of inclusion of the 3rd cumulant term Eq. (\ref%
{Airy}), labeled by Airy, are compared with the G1 and MC pdfs at
temperature $T_0=8$ K. The corresponding MATLAB function has been used to
calculate the Airy function. In general, the effect of the 3rd cumulant on
the pdf results in a slight shift of the peak position to lower energies. At 
$T_0=8$ K one can observe small oscillations in the low energy wing of the
Airy pdf. Also, the modulus of the Airy curve, labeled by $|\mathrm{Airy}|$,
is displayed.

\begin{figure}[tbp]
\includegraphics[clip=true,width=15cm,trim=1 1 1 1]{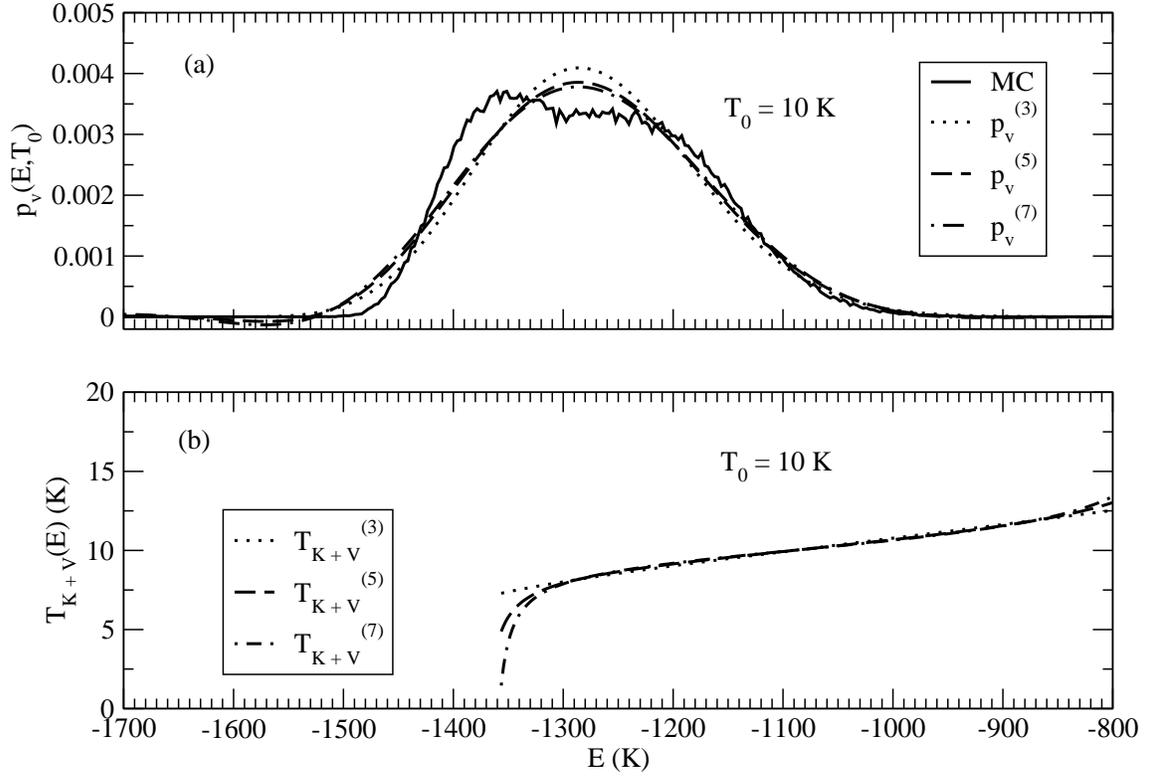}
\caption{Effects of the higher-order, $k_{max}=3,5$, and 7 cumulants on the
pdf (a) and on the statistical temperature (b) at $T_0=10$ K. Here, $p_{%
\mathcal{V}}^{(k_{max)}}$ and $T_{\mathcal{K+V}}^{(k_{max})}$ are the pdfs
and the statistical temperatures, respectively, calculated with up to the $%
k_{max}$-order cumulants included. The MC pdf shows some bimodal structure.
The system and other notations are the same as in Fig. \protect\ref{f1}.}
\label{f3}
\end{figure}

There is a hint that the MC energy distribution in Fig. \ref{f1} at $T_0=10$
K is bimodal.  It is believed that such a bimodal distribution might have a
direct connection to the solid-liquid transition in atomic clusters so that
a low-energy maximum corresponds to a solid state and a higher one to the
liquid state \cite{LW:90,WB:94,SKH:01}. In Fig. \ref{f3} (a), the MC pdf at $%
T_0=10$ K is compared to numerical estimates of the integral $p_{\mathcal{V}%
}^{(k_{max})}$ (\ref{eq33}) expressed in terms of the cumulant expansion (%
\ref{eq34d}) truncated at $k_{max}=3,5$, and 7. One can see that the
numerical results including up to the 7th-order cumulant term are not
sufficient to reproduce the hinted bimodal structure in the MC curve. It is
expected that inclusion of more cumulant terms will reproduce this
structure. Also, numerical results for the statistical temperature using the
integral representations (\ref{ST2}) are displayed in Fig. \ref{f3} (b). The
energy dependence of the temperature is seen to be close to a linear one in
the neighborhood of $\mu_{c1}=-1273$ K, as qualitatively predicted by Eq. (%
\ref{ST5}). Note that compution of the statistical temperature at the lower
energies $E<-1350$ K becomes progressively less accurate because when moving
in the low-energy region far from the central peak, where the pdf values get
smaller and, thus, relatively less accurate, this ill-defined low-energy
region turns out to make a major contribution to the convolution integral (%
\ref{eq19a}).

\begin{figure}[tbp]
\includegraphics[clip=true,width=15cm]{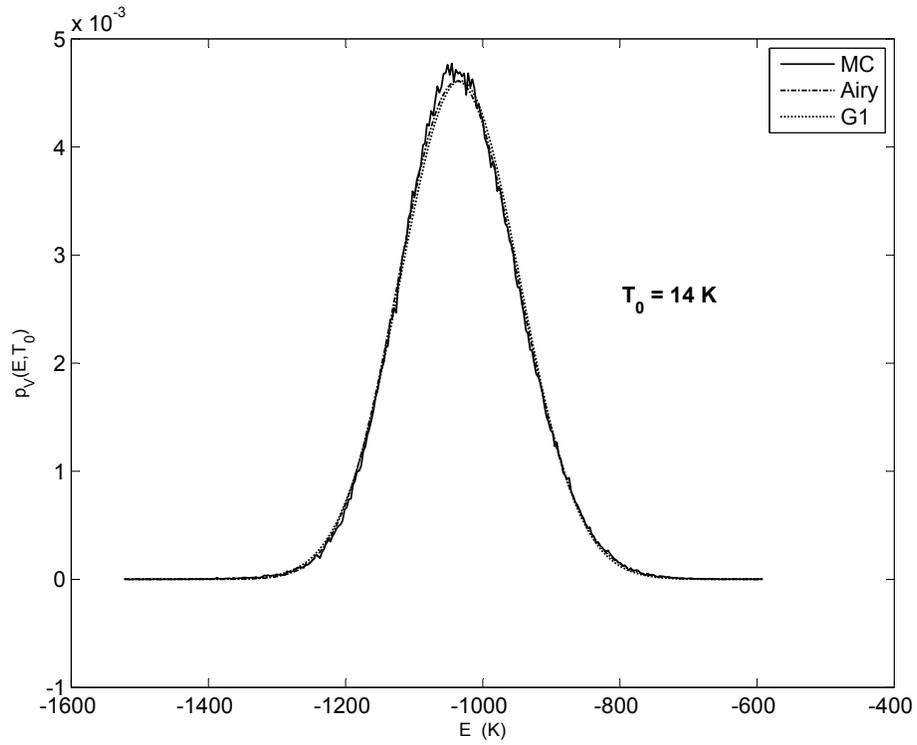}
\caption{The Airy and the Gaussian pdfs in comparison with the MC pdf at $%
T_0=14$ K. The third cumulant term has a minor effect on the pdf at $T_0=14$
K. The system and notations are the same as in Fig. \protect\ref{f2}. }
\label{f4}
\end{figure}

As temperature increases, the difference between the Airy and G1 pdfs
becomes negligible. This is demonstrated in Fig. \ref{f4} at $T_0=14$ K,
where G1 and Airy pdfs are compared with the corresponding MC results.

\begin{figure}[tbp]
\includegraphics[clip=true,width=15cm]{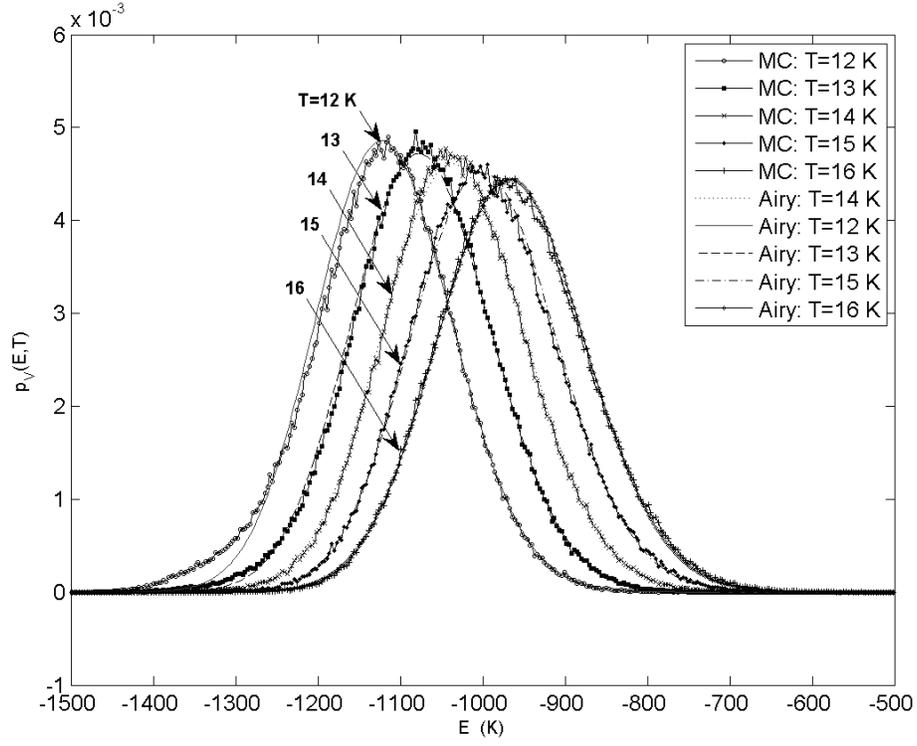}
\caption{The analytic results from formula (\protect\ref{eq36}) applied to
the Airy pdf at $T_0=14$ and continued to temperatures $T=12,13,15$, and 16
K in comparison with the corresponding MC results. The system and notations
are the same as in Fig. \protect\ref{f2}.}
\label{f5}
\end{figure}

In Fig. \ref{f5}, we demonstrate how the analytic continuation $%
T_{0}\rightarrow T$ formula (\ref{eq36}) works. With the Airy pdf taken at
temperature $T_{0}=14$ K, it is continued to $T=12,13,15$, and 16 K. For
comparison, the corresponding MC pdfs are plotted as well. The normalization
function $S^{(3)}$ has been evaluated with the help of the analytic formula (%
\ref{eq37}). Obviously, the continuation formula works better when we move
to the higher rather than lower temperatures. Moreover, the high-energy
wings of the continued pdfs are reproduced better than the low-energy ones.
Observe that the low-energy wing of the Airy pdf at $T=12$ K decays faster
than the corresponding wing of the MC pdf. The increasing difficulty of
analytic continuation to the lower temperatures can be explained by the fact
that as one can see the pdfs are shifted to the lower energies as
temperature $T$ goes down so that the overlap between the pdfs at different
temperatures becomes increasingly smaller. With $T$ decreasing, the
difference between inverse temperatures $\Delta \beta >0$ becomes bigger and
the exponential factor in (\ref{eq36}) grows exponentially. This factor
blows up the low-energy wing of the $p(E,T_{0})$ pdf and as a result we
observe that the peak maximum of $p(E,T)$ gets a shift to lower energies.
Moreover, in the regions far from the center of the $p(E,T_{0})$ pdf, errors
are expected to be dominant. As a result, multiplied by an exponentially big
factor, these errors either of systematic or statistical nature can generate
big deviations from the exact values of $p(E,T)$.

\begin{figure}[tbp]
\includegraphics[width=15cm,trim=1 1 1 1]{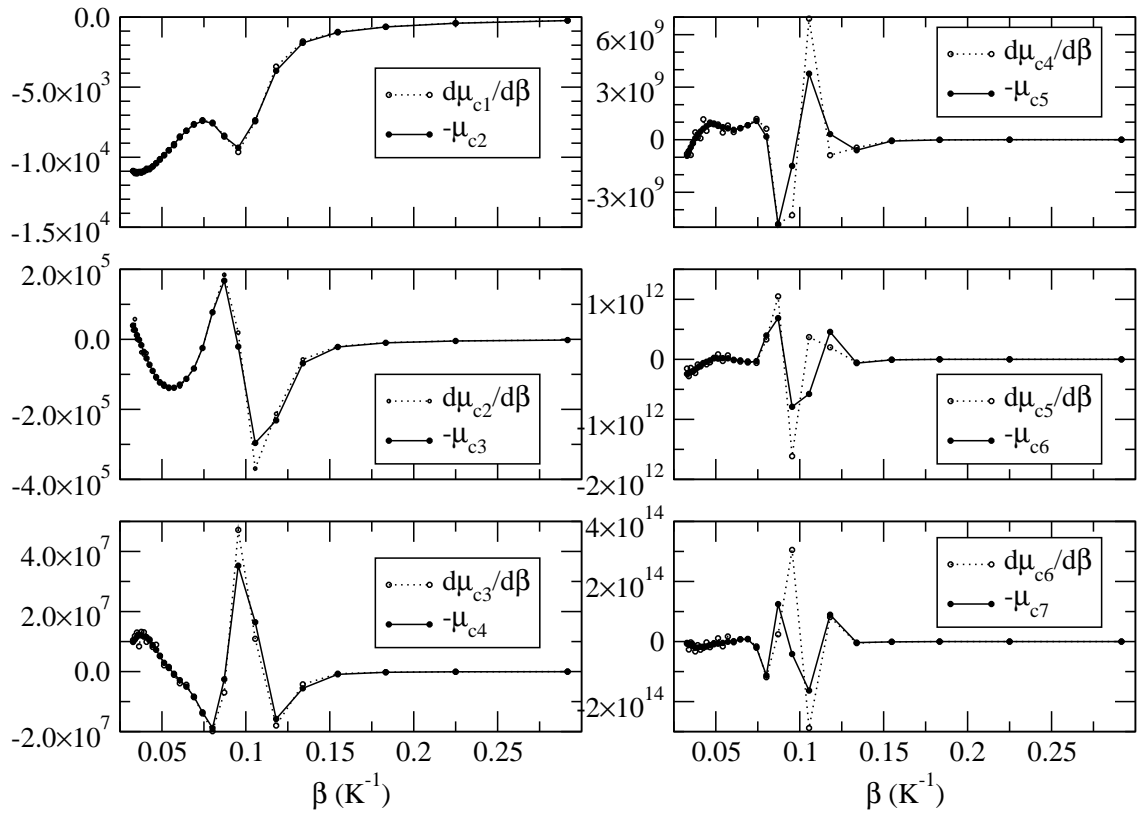}
\caption{Numerical results illustrating Theorem \protect\ref{T1}, Eq. (%
\protect\ref{eq23}) at $k=1,\ldots,6$. Numerical estimates of the
derivatives $d\protect\mu_{ck}/d\protect\beta$ as compared to $-\protect\mu%
_{c(k+1)}$. The system is the same as in Fig. \protect\ref{f1}.}
\label{f6}
\end{figure}

Fig. \ref{f6} presents numerical results supporting Theorem \ref{T1}. First,
we calculated moments up to the 7-th order on the equidistant grid with the
step $\Delta T=1$ K in the temperature range from 3 to 31 K. Evaluating
higher-order moments is computationally inexpensive since it requires only
calculation of extra powers of the potential energy. Then, with the help of
Eqs. (\ref{A5}) cumulants $\mu _{ck}$, $k=1,\ldots ,7$ can be recursively
obtained. The numerical derivatives of cumulants $d\mu _{ck}/d\beta $ are
compared with the corresponding higher-order cumulants $\mu _{c(k+1)}$ taken
with the minus sign on the inverse temperature $\beta $ scale. In general,
agreement is seen to be better for cumulants of lower orders and at higher
values of $\beta $. At smaller values of $\beta <0.05$ and in the region $%
\beta \sim 0.1$ K$^{-1}$, where the curves exhibit rapid changes, numerical
estimates of the derivatives become more scattered due to errors in the
numerical formula for the derivatives, as well as due to statistical errors
present in the MC cumulant estimates themselves. Note that when the
derivative of $\mu _{ck}$ takes a zero value at some value of $\beta _{ext}$%
, $\mu _{ck}$ as a function of $\beta $ has an extremum at this point.
According to Theorem \ref{T1}, $\mu _{c(k+1)}$ changes the sign at $\beta
_{ext}$. Fig. \ref{f6} confirms such a behavior.

\begin{figure}[tbp]
\includegraphics[clip=true,width=15cm,trim=1 1 1 1]{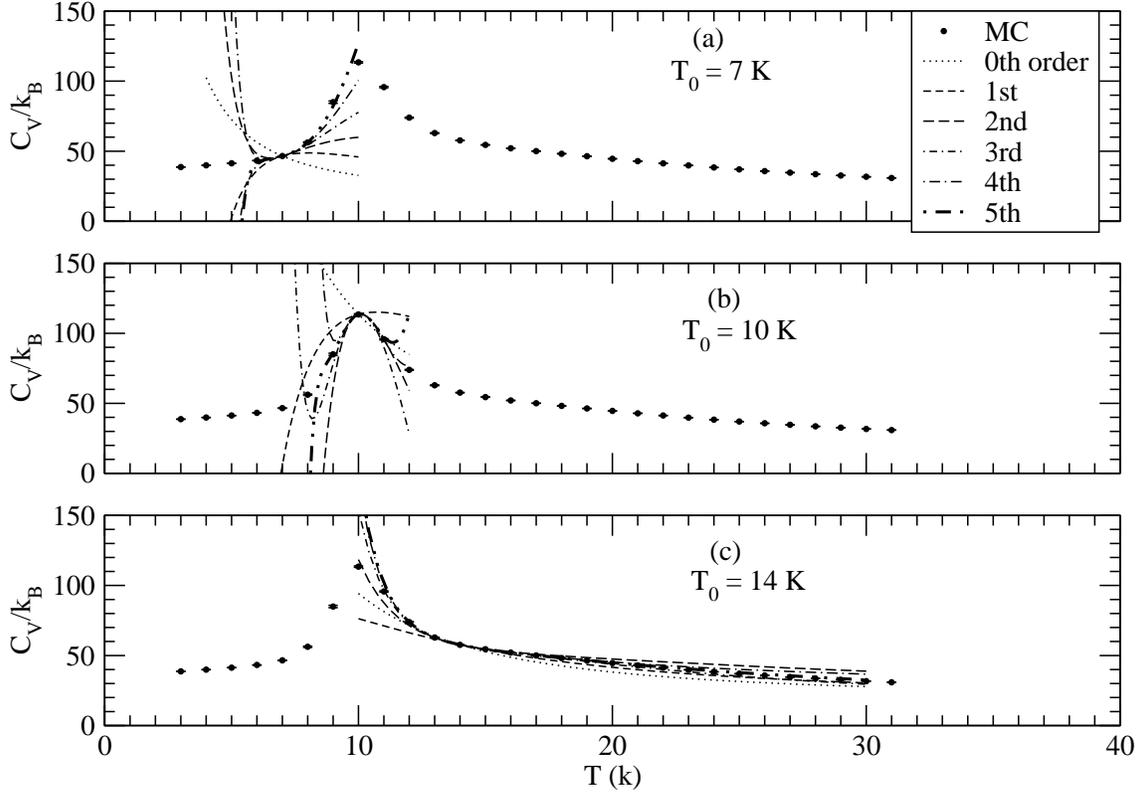}
\caption{The heat capacity $C_V$ in units of the Boltzmann constant $k_B$ as
a function of temperature $T$. The system is the same as in Fig. \protect\ref%
{f1}. The data labeled by MC have been generated with 10$^6N$ MC points in
each of 50 statistical blocks, and the error bars are at 95$\%$ confidence
level. The analytic continuation of the heat capacity value in the
neighborhoods of $T_0=7$ (a), 10 (b), and 14 K (c) are done using expansion
formula (\protect\ref{eq32}). The $k$th-order curves ($k=0,\ldots,5$) are
the results of calculation that include maximum up to the $k$th power terms
in the expansion (\protect\ref{eq32}) in $\Delta\protect\beta$ powers.}
\label{f7}
\end{figure}

In Fig. \ref{f7} (a)-(c), making use of the Taylor series expansion (\ref%
{eq32}) we present results of analytic continuation of the heat capacity
calculated at temperatures $T_{0}=7$, 10 and 14 K, respectively. For
comparison, the results of MC simulation on the equidistant grid of
temperatures with the step $\Delta T=1$ K are shown in the range from 3 to
31 K. The data have been generated with 10$^{6}N$ MC points in each
statistical block, and the error bars are at the 95\% confidence level. The
present MC data coincide within the statistical errors with the results of
independent simulations \cite{PSD:03b} obtained in the temperature range 4
to 14 K. Corresponding to the maximum power of $\Delta \beta $ terms
included in expansion (\ref{eq32}), the curves including zero, first and so
on up to the 5th-order terms are displayed. One can see that dynamics of
continuation results is generally improved with inclusion of more terms in
the expansion. Thus, we find that the results of the 5th-order curve
continuation are good in the intervals $(6,9)$, $(8.5,11),$ and $(11,31)$ K
corresponding to $T_{0}=7,\,10$ and 14 K. With increasing temperature $T_{0}$
to 14 K, the interval on which the analytic continuation works is seen to
become bigger. In the neighborhood of $T_{0}=10$ K, the heat capacity curve
achieves a maximum value. To find this peak position, we solved the
polynomial equations (\ref{peak}) for $\Delta \beta $ at $\beta _{0}=0.1$ K$%
^{-1}$ truncated at $k_{\max }=1,2,3,$ and 4. The corresponding roots were
obtained using the MATLAB function 'roots'. The results are%
\begin{equation*}
\begin{array}{ccc}
\Delta \beta _{peak}^{(k_{\max }=1)} & = & -7.905\cdot 10^{-4}\,\mathrm{K}
\\ 
\Delta \beta _{peak}^{(2)} & = & -7.697\cdot 10^{-4}\,\mathrm{K} \\ 
\Delta \beta _{peak}^{(3)} & = & -7.726\cdot 10^{-4}\,\mathrm{K} \\ 
\Delta \beta _{peak}^{(4)} & = & -7.727\cdot 10^{-4}\,\mathrm{K}%
\end{array}%
\end{equation*}%
so that the converged peak position temperature is found to be $%
T_{peak}=10.078$ K. Observe that at $T_{0}=7$ K the slope of the 0th-order
curve is negative, while the 1st and higher-order heat capacity curves show
positive slopes in qualitative agreement with the behavior of the MC curve.

\begin{figure}[tbp]
\includegraphics[clip=true,width=15cm,trim=1 1 1 1]{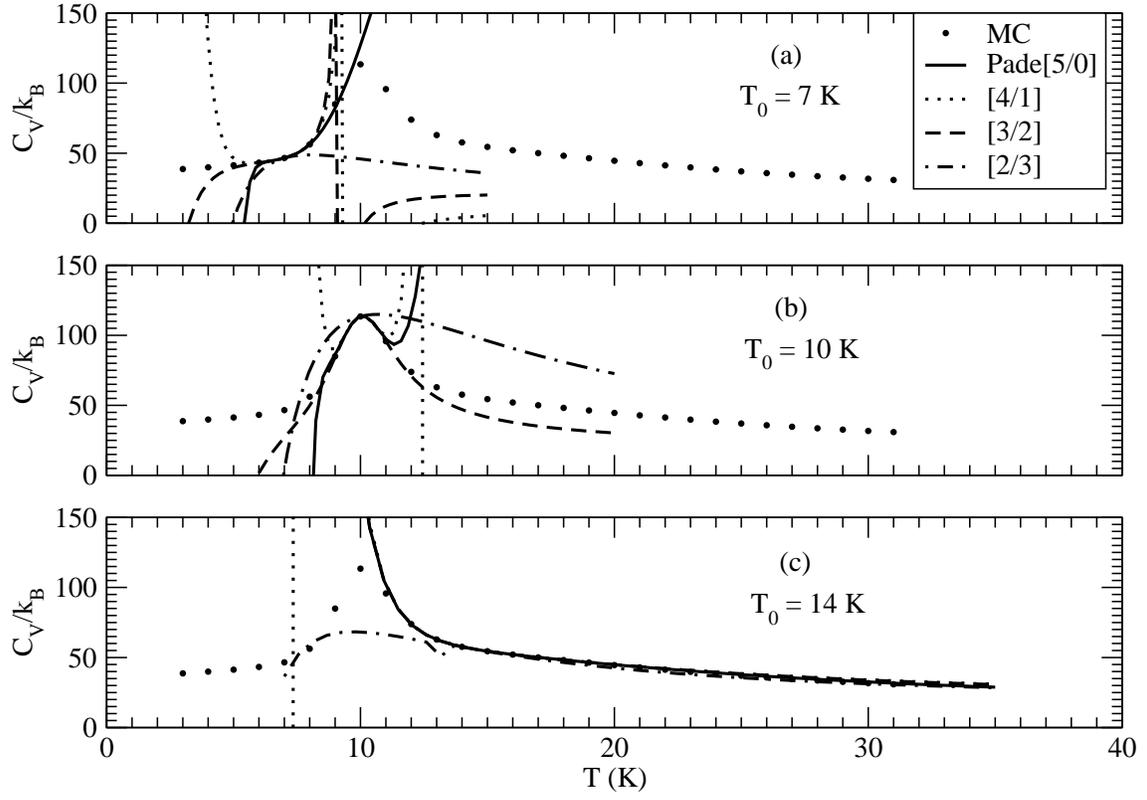}
\caption{Analytic continuation of the heat capacity by Pad\'{e} approximants
of the orders $[5/0]$, $[4,1]$, $[3,2]$, and $[2,3]$ in the neighborhoods of 
$T_0=7$ (a), 10 (b), and 14 K (c). The system and other labels are the same
as in Fig. \protect\ref{f7}.}
\label{f8}
\end{figure}

The Pad\'{e} approximants often give better approximation of the function
than truncating its Taylor series, and it may still work where the Taylor
series does not converge \cite{BGM:96}. In Fig. \ref{f8} (a)-(c), we compare
the MC results with the Pad\'{e} approximants of various orders $\left[ 5/0%
\right] ,$ $\left[ 4/1\right] ,$ $\left[ 3/2\right] $, and $\left[ 2/3\right]
$ calculated in the neighborhoods of temperatures $T_{0}=7$, 10, and 14 K
respectively. Notice that the Pad\'{e} approximant $\left[ 5/0\right] $
coincides with the Taylor series truncated at the 5th order. In general,
depending on $T_{0}$, the best agreement is seen either for $[5/0]$ or $%
[3,2] $ Pad\'{e} approximants. Thus, observe that at $T_{0}=10$ K the Pad%
\'{e} approximant $[3/2]$ shows a slightly better behavior than that of $%
[5/0]$.

\begin{figure}[tbp]
\includegraphics[clip=true,width=15cm,trim=1 1 1 1]{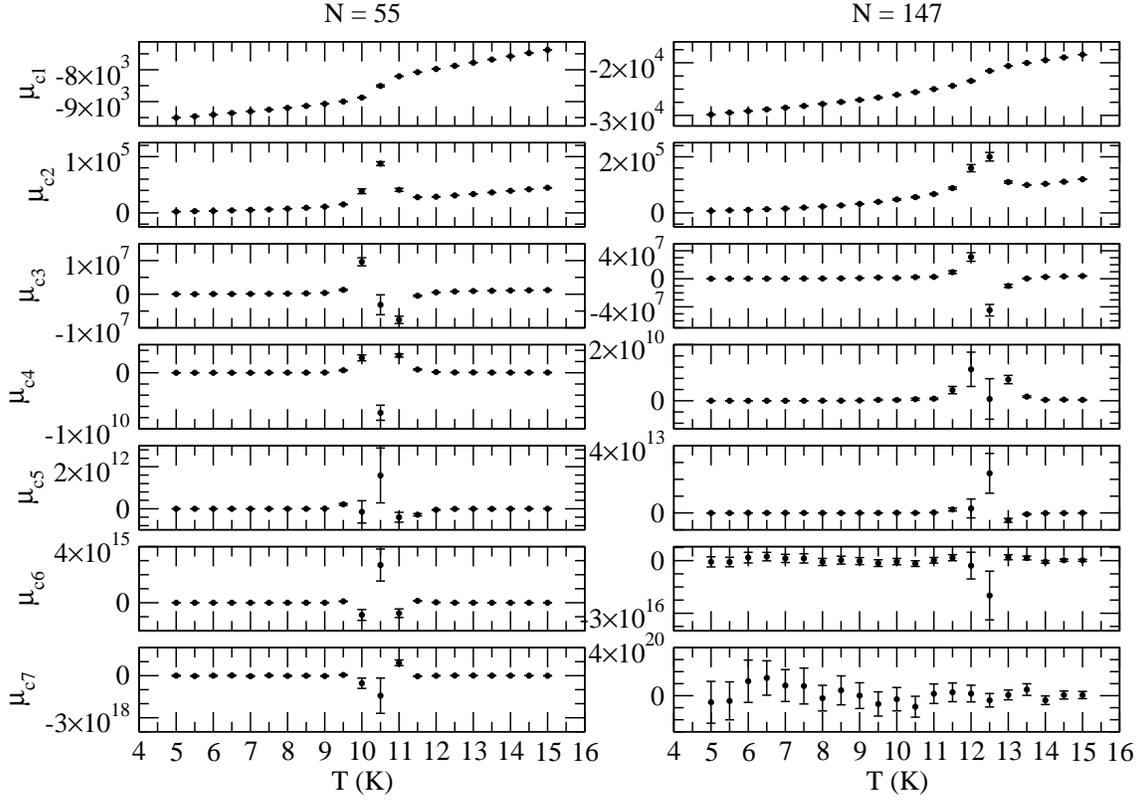}
\caption{The temperature dependence of cumulants $\mu_{ck}$, $k=1,\ldots,7$, calculated for the system of $N=55$ (left) and 147 (right panel) neon particles. The MC data have been generated with 10$^6N$ and $5\cdot 10^5N$ MC points in the left and right panels, respectively, in each of 50 statistical blocks. The error bars are at 95$\%$ confidence level.}
\label{f9}
\end{figure}

\begin{figure}[tbp]
\includegraphics[clip=true,width=15cm,trim=1 1 1 1]{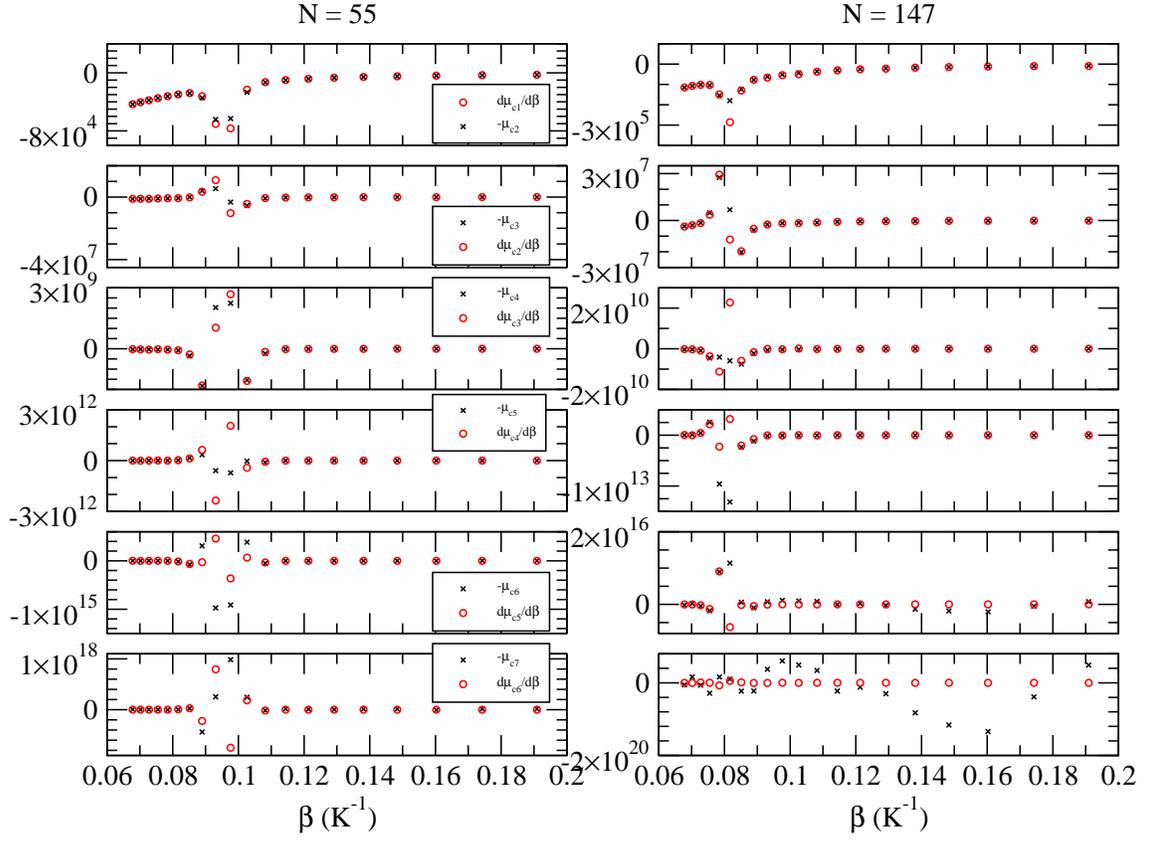}
\caption{Numerical estimates of the
derivatives $d\mu_{ck}/d\beta$, $k=1,\ldots,6$ (red circles) as compared to $-\mu_{c(k+1)}$ (black crosses). The system and statistics are the same as in Fig.~\ref{f9}.}
\label{f10}
\end{figure}

\begin{figure}[tbp]
\includegraphics[clip=true,width=15cm,trim=1 1 1 1]{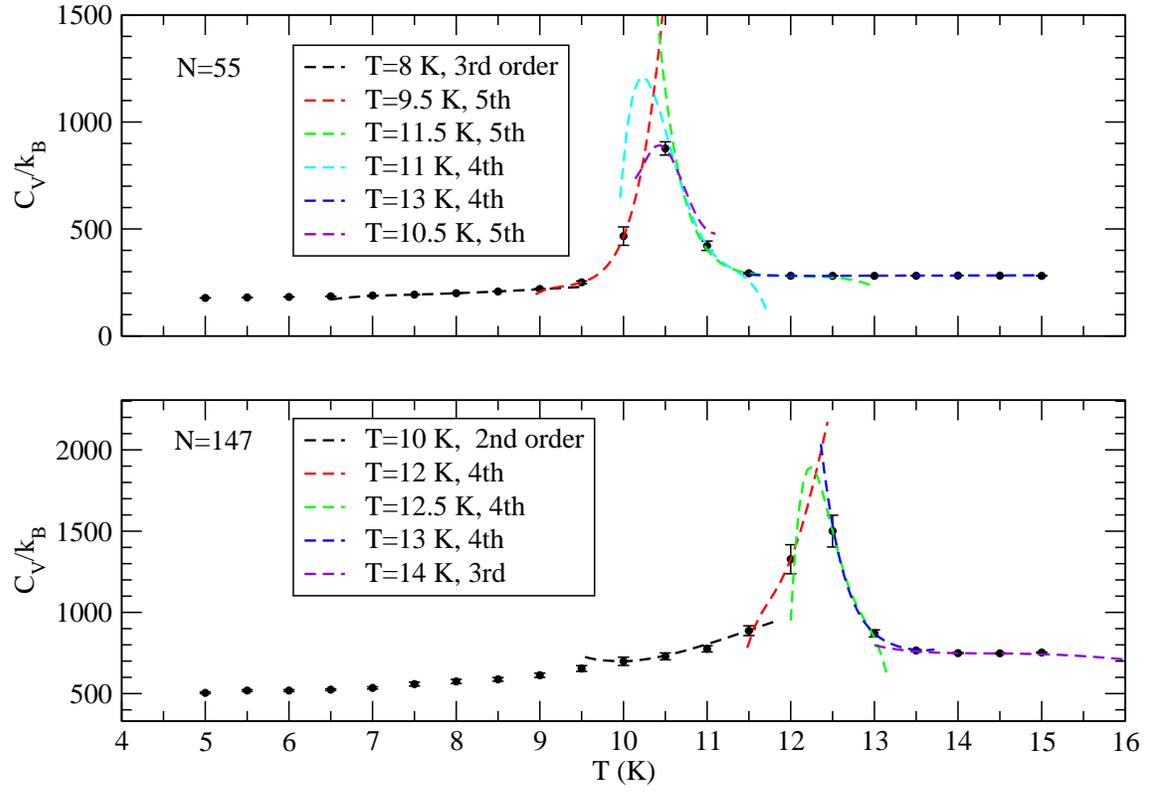}
\caption{The results of analytic continuation obtained with the help of formula (\protect\ref{eq32}) for the system of $N=55$ (upper) and 147 (lower panel) neon particles. Notations are the same as in Fig. \ref{f7}. Statistics of the MC moves are the same as in Fig. \ref{f9}.}
\label{f11}
\end{figure}

In Figs. \ref{f9}-\ref{f11}, the corresponding results for the cumulants, 
their derivatives and the heat capacity curves are displayed for $N=55$ and 147 neon 
particles. 
The confining radius is set to be $R_{c}=0.85\sigma _{LJ}N^{1/3}$.
Notice that the MC data have been generated with $N_{bl}=10^{6}N$ MC moves for the system of $N=55$ particles and 
with $N_{bl}=5\cdot 10^5N$  MC points for $N=147$ particles in each of 50 statistical blocks.
The size of the error bars obtained for the cumulants in Fig.~\ref{f9}  tends to become bigger 
for the cumulants of higher  orders, but decrease when temperature goes up, 
except for the region where the heat capacity reveals a peak. 
In Fig.~\ref{f10}, the derivatives $d\mu_{ck}/d\beta$, $k=1,\ldots,6$ are compared to the corresponding cumulants $-\mu_{c(k+1)}$. The agreement is quite good, 
additionally supporting Theorem \ref{T1}. Deviations between $d\mu_{ck}/d\beta$ and $-\mu_{c(k+1)}$ are most pronounced in the regions of cumulant's rapid change and these deviations are caused either by the statistical or systematic errors  in numerical estimates of the derivatives and the corresponding cumulants.
Thus, one can see that the 7th-order cumulant for $N=147$ particles, shown in the right panel of Fig.~\ref{f9}, is poorly converged at current value of $N_{bl}$ since it  is almost totally in error at $T\le 12$ K. As expected, this results in a  poor agreement observed  in the right panel of Fig.~\ref{f10}  between the derivative $d\mu_{c6}/d\beta$ and the cumulant $-\mu_{c7}$. Fig.~\ref{f11} demonstrates the results of analytic continuation obtained for the heat capacity $C_{V}$ using formula (\ref{eq32}). From analytic continuation curves we  get the following estimates for the peak positions $T_{peak}=10.44$ and $12.25$ K for the system of $N=55$  and 147 particles, respectively.

Finally, let us consider the convergence issue, that is, how errors present in the MC estimates may affect the results of continuation. The MC estimates for cumulants can be represented as $\mu_{ck}=\mu_{ck}^{exact}+\varepsilon_{ck}$, where $\mu_{ck}^{exact}$ is an exact value of the $k$th-order cumulant and $\varepsilon_{ck}$ is an error in its estimate. We assume that cumulant estimates have been generated at a fixed number of MC moves. The longer MC moves are generated the less errors one gets in $\mu_{ck}$'s. Notice that we cannot directly compare the sizes of errors in cumulants of different orders since they have {\it different} physical dimensions: $[\mu_{ck}]=[{\rm Energy}]^k$. In the continuation formula (\ref{eq32}), the $k$th-order cumulant enters multiplied by the $k$th power of a small expansion parameter $\Delta\beta$, so that  the continued 2nd-order cumulant can be written as
\begin{eqnarray}
\mu_{c2}(\beta)&=&\mu_{c2}^{exact}(\beta)+\varepsilon_{c2}^{tot}(\beta) ,\nonumber\\
\mu_{c2}^{exact}(\beta)&=&\mu_{c2}^{exact}(\beta_0)-\dfrac{\mu_{c3}^{exact}(\beta_0)}{1!}\Delta\beta+\dfrac{\mu_{c4}^{exact}(\beta_0)}{2!}(\Delta\beta)^2+\cdots, \nonumber\\
\varepsilon_{c2}^{tot}(\beta)&=&\varepsilon_{c2}(\beta_0)-\dfrac{\varepsilon_{c3}(\beta_0)}{1!}\Delta\beta+\dfrac{\varepsilon_{c4}(\beta_0)}{2!}(\Delta\beta)^2+\cdots
\end{eqnarray}
One can see that the total error $\varepsilon_{c2}^{tot}(\beta)$ in the continued 2nd-order cumulant is defined by the sum of errors in the 2nd- and higher-order cumulants at $\beta_0$, contributions from the higher-order cumulant's errors are being multiplied by the powers of the expansion parameter $\Delta\beta$. With the growth of $\Delta\beta$, the contribution of the 3rd- and higher-order cumulant's errors to $\varepsilon_{c2}^{tot}(\beta)$ increases and at some value, which we call a radius of convergence $\Delta\beta_{conv}$, its contribution  becomes comparable to the size of the 2nd-order cumulant's error $\varepsilon_{c2}(\beta_0)$. We do not know exactly errors in cumulants (errors are random numbers) but their size can be estimated by 2$\sigma_{ck}$'s, by the two standard deviations in a usual way (by error bars in Fig. \ref{f9}). Thus, one can estimate the  raduis of convergence $\Delta\beta_{conv}^{(k\to2)}$ for the $k$th-order cumulant ($k>2$) by the expression
\begin{equation}
|\Delta\beta_{conv}^{(k\to2)}|=\left((k-2)!\dfrac{\sigma_{c2}}{\sigma_{ck}}\right)^{1/(k-2)}
\end{equation}
Using the relationship between inverse and direct temperature scales $\Delta\beta=\beta-\beta_0=1/T-1/T_0\approx -\Delta T/T_0^2$, where $\Delta T=T-T_0$, one finds the corresponding radius of convergence in the temperature scale
\begin{equation}
\Delta T_{conv}^{(k\to2)}=T_0^2\left((k-2)!\dfrac{\sigma_{c2}}{\sigma_{ck}}\right)^{1/(k-2)}
\label{Tconv}
\end{equation}
It roughly defines the temperature range within which the errors present in the $k$th-order cumulant will produce the same order errors as in the $\mu_{c2}(\beta_0)$.

The $2\sigma_{ck}$'s obtained for the cluster of $N=55$ neon particles at low $T_0=5$ and 15 K and high temperatures $T_0=100$ and 300 K are summarized in Table \ref{Tab1}. Observe how the magnitude of errors in cumulants increases, roughly by an order, as the number of MC moves specified by the parameter $f_{MC}$ decreases by two orders at $T_0=100$ K. 
\begin{table}
\caption{\label{Tab1}The two standard deviations $2\sigma_{ck}$ ($k=2,\ldots,7$) calculated for the $k$th-order cumulant at low $T_0=5$, 15 (the size of error bars in Fig. \ref{f9}) and at high temperatures $T_0=100$ and 300 K using 50 statistical blocks. The number of Monte Carlo moves in a single statistical block is $N_{MC}=f_{MC}N$, where $N=55$ is the number of neon particles in the LJ cluster.}
\begin{ruledtabular}
\begin{tabular}{llllllll}
  $T_0$ (K) & $f_{MC}$ & $2\sigma_{c2}$ (K$^2$) & $2\sigma_{c3}$ (K$^3$) & $2\sigma_{c4}$ (K$^4$) & $2\sigma_{c5}$ (K$^5$) & $2\sigma_{c6}$ (K$^6$) & $2\sigma_{c7}$ (K$^7$)\\
\hline
  5 & 10$^6$ & $3.0\cdot10^0$ & $2.5\cdot 10^2$ & $2.2\cdot 10^4$ & $1.2\cdot 10^8$ & $2.2\cdot 10^{12}$ & $4.2\cdot 10^{16}$\\
  15 & 10$^6$ & $1.6\cdot 10^2$ & $4.6\cdot 10^4$ & $1.9\cdot 10^7$ & $7.7\cdot 10^9$ & $4.8\cdot 10^{12}$ & $7.0\cdot 10^{15}$\\
  100 & 10$^6$ & $1.0\cdot 10^2$ & $4.0\cdot 10^4$ & $3.5\cdot 10^7$ & $3.2\cdot 10^{10}$ & $3.2\cdot 10^{13}$ & $5.0\cdot 10^{16}$\\
100 & 10$^5$ & $2.5\cdot 10^2$ & $1.6\cdot 10^5$ & $1.1\cdot 10^8$ & $1.2\cdot 10^{11}$ & $1.8\cdot 10^{14}$ & $3.7\cdot 10^{17}$\\
100 & 10$^4$ & $6.7\cdot 10^2$ & $4.7\cdot 10^5$ & $2.9\cdot 10^8$ & $2.5\cdot 10^{11}$ & $2.3\cdot 10^{14}$ & $2.1\cdot 10^{17}$\\
300 & 10$^6$ & $3.9\cdot 10^2$ & $6.5\cdot 10^5$ & $1.7\cdot 10^9$ & $5.2\cdot 10^{12}$ & $2.1\cdot 10^{16}$ & $1.1\cdot 10^{20}$\\
\end{tabular}
\end{ruledtabular}
\end{table}

\begin{table}
\caption{\label{Tab2} The radius of convergence for the $k$th-order cumulant $\Delta T_{conv}^{(k\to2)}$ calculated with the help of Eq. (\ref{Tconv}) and the data from Table \ref{Tab1}. Other notations are the same as in Table \ref{Tab1}.}
\begin{ruledtabular}
\begin{tabular}{lllllll}
  $T_0$ (K) & $f_{MC}$ & $\Delta T_{conv}^{(3\to2)}$ (K) & $\Delta T_{conv}^{(4\to2)}$ (K) & $\Delta T_{conv}^{(5\to2)}$ (K) & $\Delta T_{conv}^{(6\to2)}$ (K) & $\Delta T_{conv}^{(7\to2)}$ (K)\\
\hline
  5 & 10$^6$ & 0.3 & 0.4 & 0.6 & 0.06 & 0.04 \\
 15 & 10$^6$ & 0.8 & 0.9 & 1.1 & 1.2 & 1.2\\
  100 & 10$^6$ & 25 & 24 & 26 & 29 & 30\\
100 & 10$^5$ & 16 & 21 & 23 & 24 & 24\\
100 & 10$^4$ & 14 & 21 & 25 & 29 & 33\\
300 & 10$^6$ & 54 & 61 & 69 & 74 & 76\\
\end{tabular}
\end{ruledtabular}
\end{table}

Making use of the data in Table \ref{Tab1}, we can estimate radii of convergence for the higher-order (3-7) cumulants with the help of Eq. (\ref{Tconv}); the results of evaluation are summarized in Table \ref{Tab2}. Observe that at $T_0=5$ K, the higher-order, 6th and 7th-order, cumulants appear to be non-converged since their radii of convergence are by an order of magnitude smaller than the values obtained for the 3rd-, 4th-, and 5th-order cumulants. At higher temperatures shown in Table \ref{Tab2}, all the cumulants seem to be well converged; the smallest radius of convergence is generally found for the 3rd-order cumulant's contribution. As an empirical rule, we find that the radius of convergence is roughly proportional to the temperature $T_0$. Notice that at $T_0=100$ K, when the number of MC moves drops down by two orders, the radius of convergence $\Delta T_{conv}^{(3\to2)}$ is  reduced by about two times only.

\section{Concluding remarks}

\label{CR}

In obtaining reliable estimates of thermal averages in realistic,
multidimensional, classical or quantum many- or few-body systems, the MC
simulation is often the only way of getting right answers. The simulated
averages depend on thermodynamic parameters, such as temperature, volume
etc., and/or interaction parameters between the particles. Therefore, in
order to avoid many time-consuming runs of the MC codes at various
parameters, it is important to develop robust analytic techniques that allow
us to continue the MC data in the neighborhood of a set of prescribed
parameter values. The key finding of this work is a simple relationship Eq. (%
\ref{eq23}) between derivatives and the higher-order cumulants. Since many
important thermodynamic quantities, such as energy and heat capacity, can be
expressed in terms of cumulants, this theorem provides an analytic tool or bridge to
continue MC data in the neighborhood of a control parameter value, say, $%
\beta_0$ or to fill a gap by constructing an analytic bridge in between of two neighboring parameters $\beta_0$ and $\beta_1$. 

To find an optimal numerical scheme to evaluate the cumulants up to the $k_{max}$th-order (in our examples $k_{max}=7$), it is important to estimate the amount of numerical work required. A reasonable estimate of this time is the number of potential function calls $N_{call}$ required to compute a thermodynamic quantity, e.g., the heat capacity. In a single MC move, the system moves to a new position  and one has to calculate the potential function $V_i=V(X_i)$ one time  at a new position $X_i$ so that $N_{call}=N_{bl}$ where $N_{bl}$ is a total number of MC moves in a single statistical block. To calculate cumulants up to the $k_{max}$th-order, one has to compute additionally $k_{max}-1$ powers $V_i^k$, $k=2,\ldots,k_{max}$ or perform $k_{max}-1$ multiplications of the known potential function value $V_i$. These multiplications are computationally cheap and do not depend on the size of the system. Therefore, as estimated the amount of numerical work to evaluate the higher-order cumulants will be practically the same as in the standard parallel tempering scheme which requires evaluation of only $V_i$ and $V_i^2$ values.

By Eqs. (\ref{eq33})-(\ref{eq34c}), the energy pdfs can also be expressed in
terms of cumulants. Truncating the cumulant expansion at $k_{max}$, we have
derived analytic expressions (\ref{pdf1}), (\ref{eq35}), and (\ref{Airy})
for the pdfs at $k_{max}=1,2,$ and 3 respectively. The higher-order energy
pdfs truncated at $k_{max}\ge 4$ can be obtained either by evaluating the $%
\tau$-integral numerically or by applying an appropriate (saddle-point)
asymptotic method.

Generalizations to a multi-parameter classic system are rather
straightforward and can be expressed by the Theorem \ref{T2}. Using this
theorem, one can easily develop similar analytic continuation formulas, such
as Eqs. (\ref{6.3}), in terms of the higher-order multi-variate cumulants in
the multi-parametric space. The path integral Feynman-Kac representation of
the density matrix is a very useful formulation of the quantum statistical
equilibrium operator since it basically reduces the problem of analytic
continuation in quantum case to the corresponding problem in
multi-parametric classical systems. Of course, technically this reduction
can be done if the original infinite-dimensional integration over the path
variables can be replaced somehow by a finite-dimensional one. We considered
several possible scenarios, having different convergence properties with
respect to the number of path variables, of such infinite-to-finite
dimensional integral replacements in Section \ref{QM}.

Numerical testing of Theorem \ref{T1}, analytic continuation formulas for
the energy pdfs and heat capacity curves has been exemplified by an LJ
classical cluster sytem in Section \ref{RD}. By now, usefulness of analytic
continuation formulas in multi-parametric classical and quantum systems have
not yet been tested numerically. Further numerical investigations of
interesting multi-parameter systems are of special interest. The results of
such investigations when available will be published elsewhere.

\begin{acknowledgments}
This work was supported by NRF (National Honor Scientist Program:
2010-0020414, WCU: R32-2008-000-10180-0) and KISTI (KSC-2011-G3-02).
\end{acknowledgments}

\appendix

\section{Moments versus Cumulants}

\label{App1}

Relationships between univariate cumulants and moments can be derived from
the moment-generating function \cite{Ku:62} 
\begin{eqnarray}
M(t) &=&\left\langle \exp \left( tH\right) \right\rangle
=1+\sum_{k=1}^{\infty }\dfrac{t^{k}}{k!}\mu _{k}  \notag \\
&=&\exp \left[ \sum_{k=1}\dfrac{t^{k}}{k!}\mu _{ck}\right]  \label{A0}
\end{eqnarray}%
where $\mu _{k}\equiv \left\langle H^{k}\right\rangle $ are called moments.
Assuming the average of 1 to be nonzero, we conclude that $\ln \left\langle
\exp \left( tH\right) \right\rangle $ is an analytic function of $t$, in a
vicinity of zero. Therefore, it has a Taylor expansion with respect to $t$.
This expansion is called \textit{cumulant expansion}; the coefficients $%
\mu_{ck}$ of the expansion are called \textit{cumulants}. Thus, cumulants in
terms of moments are expressible by the formula 
\begin{equation}
\mu _{ck}=\sum_{r=1}^{k}\dfrac{(-1)^{r+1}}{r}\sum_{p_{1},\ldots ,p_{r}\in
C_{kr}}\dfrac{k!}{p_{1}!\cdots p_{r}!}\mu _{p_{1}}\cdots \mu _{p_{r}}
\end{equation}%
where the summation of the integer indices $p_{1},\cdots ,p_{r}\geq 1$ is
restricted by the relation 
\begin{equation*}
C_{kr}=\left\{ (p_{1},\cdots ,p_{r}):\;\sum_{j=1}^{r}p_{j}=k\right\}
\end{equation*}%
In order to calculate the $k$-th order cumulant one needs moments up to the $%
k$-th order.

On the other hand, moments can be expressed via cumulants by a similar
formula 
\begin{eqnarray}
\mu_k=\sum_{r=1}^k\dfrac 1{r!}\sum_{\substack{ p_1,\ldots,p_r\in C_{kr}}} 
\dfrac{k!}{p_1!\cdots p_r!}\mu_{cp_1}\cdots\mu_{cp_r}  \label{A3}
\end{eqnarray}
The $k$th moment $\mu_k$ is a $k$th-degree polynomial in the first $k$
cumulants. These polynomials have a remarkable combinatorial interpretation:
the coefficients count certain partitions of sets. A general form of these
polynomials is 
\begin{eqnarray}
\mu_k=\sum_{\pi}\prod_{S\in\pi}\mu_{c|S|}  \label{A4}
\end{eqnarray}
where $\pi$ runs through the list of all partitions of a set of size $k$; "$%
S\in\pi$" means $S$ is one of the "blocks" into which the set is
partitioned; and $|S|$ is the size of the set $S$. To better understand how
Eq. (\ref{A4}) corresponds to (\ref{A3}) let us consider all possible
partitions of the set of natural numbers $\{1,2,\ldots,k\}$. We can divide
all possible partitions $\{\pi\}$ into disjoint groups $\{\pi\}=\{\pi_1,%
\ldots,\pi_k\}$ such that the number of blocks in a particular partition $%
\pi_r$, $r=1,\ldots,k$ is equal to $r$. For example, in case of $k=3$ we
obtain the set of all possible partitions classified as $\{\pi_1,\pi_2,\pi_3%
\}$. Here, $\pi_1$ includes an "improper" partition $(123)$. The size of the 
$(123)$ partition is three. $\pi_2$ corresponds to partitions into two
blocks: $(12)3$, $(13)2$, $(23)1$, with sizes of blocks being two and one. $%
\pi_3$ splits the set into three blocks $(1)(2)(3)$, each block of size one.
Round parenthesis show how the set is partitioned. The total number of
partitions of a $k$-element set is the Bell number $B_k$: $B_0=1$, $B_1=1$, $%
B_2=2$, $B_3=5$. Bell numbers satisfy the recursion $B_{k+1}=\sum_{n=0}^k 
\binom{k}{n} B_n$, where $\binom{k}{n}$ is the binomial coefficient.

If $S_{1},\ldots ,S_{r}\in \pi _{r}$ are $r$ blocks into which the set can
be divided, then Eq. (\ref{A4}) can be rewritten as 
\begin{equation}
\mu _{k}=\sum_{r=1}^{k}\sum_{S_{1},\ldots ,S_{r}\in \pi _{r}}\mu
_{c|S_{1}|}\cdots \mu _{c|S_{r}|}  \label{A4a}
\end{equation}%
where inner summation runs over all possible non-empty, disjoint blocks.
Then, each $r$-term in Eq. (\ref{A3}) corresponds to the $r$-term in (\ref%
{A4a}): 
\begin{equation}
\dfrac{1}{r!}\sum_{\substack{ p_{1},\ldots ,p_{r}\in C_{kr}}}\dfrac{k!}{%
p_{1}!\cdots p_{r}!}\mu _{cp_{1}}\cdots \mu _{cp_{r}}=\sum_{S_{1},\ldots
,S_{r}\in \pi _{r}}\mu _{c|S_{1}|}\cdots \mu _{c|S_{r}|}  \label{A4b}
\end{equation}%
Indeed, the multinomial coefficient 
\begin{equation}
\left( 
\begin{array}{ccc}
& k &  \\ 
p_{1} & \cdots & p_{r}%
\end{array}%
\right) =\dfrac{k!}{p_{1}!\cdots p_{r}!}  \label{mCoef}
\end{equation}%
in the left-hand-side is the number of ways of grouping $k$ objects
(numbers) into $r$ groups (blocks) of sizes $p_{1},\ldots ,p_{r}$, when the
order within each group does not matter. The summation over blocks in the
right-hand side can be carried out into two steps. First, the summation over
all possible block's sizes $|S_{1}|=p_{1},\ldots ,|S_{r}|=p_{r}$ such that $%
p_{1}\leq \cdots \leq p_{r}$ and then summing up over blocks at fixed sizes
of blocks 
\begin{equation*}
\sum_{S_{1},\ldots ,S_{r}\in \pi _{r}}\ldots =\sum_{\substack{ p_{1}\leq
\cdots \leq p_{r}  \\ p_{1}+\cdots +p_{r}=k}}\sum_{|S_{1}|=p_{1},\ldots
,|S_{r}|=p_{r}}\ldots
\end{equation*}%
Obviously, the inner summation will result in the same multinomial
coefficient (\ref{mCoef}) as in the left-hand side of (\ref{A4b}). Finally,
notice that the summed function $\mu _{cp_{1}}\cdots \mu _{cp_{r}}$ and the
multinomial coefficient are totally symmetric functions with respect to
permutations of $p_{1},\ldots ,p_{r}$ indexes and $r!$ is the number of
their permutations. This symmetry results in 
\begin{equation*}
\dfrac{1}{r!}\sum_{\substack{ p_{1},\cdots ,p_{r}  \\ p_{1}+\cdots +p_{r}=k}}%
\ldots =\sum_{\substack{ p_{1}\leq \cdots \leq p_{r}  \\ p_{1}+\cdots
+p_{r}=k }}\ldots
\end{equation*}%
This proves Eq. (\ref{A4b}).

Using Eq. (\ref{A3}) or (\ref{A4a}), one obtains the following expressions
for the first seven moments in terms of cumulants 
\begin{eqnarray}
\mu_1&=&\mu_{c1},  \notag \\
\mu_2&=&\mu_{c2}+\mu_{c1}^2,  \notag \\
\mu_3&=&\mu_{c3}+3\mu_{c2}\mu_{c1}+\mu_{c1}^3,  \notag \\
\mu_4&=&\mu_{c4}+4\mu_{c3}\mu_{c1}+3\mu_{c2}^2+
6\mu_{c2}\mu_{c1}^2+\mu_{c1}^4,  \notag \\
\mu_5&=&\mu_{c5}+5\mu_{c4}\mu_{c1}+10\mu_{c2}\mu_{c3}+
10\mu_{c3}\mu_{c1}^2+15\mu_{c2}^2\mu_{c1}+ 10\mu_{c2}\mu_{c1}^3+\mu_{c1}^5, 
\notag \\
\mu_6&=&\mu_{c6}+6\mu_{c5}\mu_{c1}+15\mu_{c2}\mu_{c4}+
10\mu_{c3}^2+15\mu_{c4}\mu_{c1}^2  \notag \\
&+&60\mu_{c3}\mu_{c2}\mu_{c1}+15\mu_{c2}^3
+20\mu_{c3}\mu_{c1}^3+45\mu_{c2}^2\mu_{c1}^2+\mu_{c1}^6,  \notag \\
\mu_7&=&\mu_{c7}+7\mu_{c6}\mu_{c1}+21\mu_{c2}\mu_{c5}+
35\mu_{c3}\mu_{c4}+21\mu_{c5}\mu_{c1}^2  \notag \\
&+&105\mu_{c4}\mu_{c2}\mu_{c1}+70\mu_{c3}^2\mu_{c1}
+105\mu_{c3}\mu_{c2}^2+35\mu_{c4}^2\mu_{c1}^3  \notag \\
&+&210\mu_{c2}\mu_{c3}\mu_{c1}^2+105\mu_{c2}^3\mu_{c1}
+35\mu_{c3}\mu_{c1}^4+105\mu_{c2}^2\mu_{c1}^3+21\mu_{c2}\mu_{c1}^5
+\mu_{c1}^7  \label{A5}
\end{eqnarray}
These equations relating the cumulants and the moments can also be
interpreted as recurrence relations that allow the expression of the
higher-order cumulants in terms of lower-order ones.

\section{Proof of Eq. (\ref{eq23})}

\label{App2}

The proof is by induction. At $k=1$, the statement of theorem follows
directly from Eq. (\ref{eq22}) and definitions of the first two cumulants $%
\mu _{c1}=\mu _{1}$ and $\mu _{c2}=\mu _{2}-\mu _{1}^{2}$. Let us assume
that Eq. (\ref{eq23}) is valid at $k=1,\ldots ,p$; we have to prove its
validity at $k=p+1$. Making use of combinatorial representation (\ref{A4a}),
the $(p+1)$-st moment can be written as \cite{KFD:10} 
\begin{equation}
\mu _{p+1}=\sum\limits_{r=1}^{p+1}\sum\limits_{S_{1},\ldots ,S_{r}}\mu
_{c|S_{1}|}\ldots \mu _{c|S_{r}|}  \label{eq24}
\end{equation}%
where $S_{1},\ldots ,S_{r}$ denote a partition of a set of natural numbers $%
\left\{ 1,\ldots ,p+1\right\} $. If blocks $S_{1},\ldots ,S_{r}$ contain,
respectively, $k_{1},\ldots ,k_{r}$ elements [numbers] such that $%
k_{1}+\ldots +k_{r}=p+1$, then $\mu _{c|S_{1}|}=\mu _{ck_{1}},\ldots ,\mu
_{c|S_{r}|}=\mu _{ck_{r}}$. The term in (\ref{eq24}) for $r=1$ corresponds
to $\mu _{c(p+1)}$, so that Eq. (\ref{eq24}) can be rewritten as%
\begin{equation}
\mu _{c(p+1)}=\mu _{p+1}-\sum\limits_{r=2}^{p+1}\sum\limits_{S_{1},\ldots
,S_{r}}\mu _{c|S_{1}|}\ldots \mu _{c|S_{r}|}  \label{eq25}
\end{equation}%
where the summation is performed over the "proper partitions" when $r\geq 2$%
. Observe that the sizes $k_{1},\ldots ,k_{r}$ of "proper partitions" in (%
\ref{eq25}) cannot be bigger than $p$. Differentiating (\ref{eq25}) with
respect to $\beta $, we get 
\begin{eqnarray}
\dfrac{d\mu _{c(p+1)}}{d\beta } &=&-\mu _{p+2}+\mu _{p+1}\mu _{1}  \notag \\
&-&\sum\limits_{r=2}^{p+1}\sum\limits_{S_{1},\ldots ,S_{r}}\dfrac{\mu
_{c|S_{1}|}}{d\beta }\ldots \mu _{c|S_{r}|}-\cdots  \notag \\
&-&\sum\limits_{r=2}^{p+1}\sum\limits_{S_{1},\ldots ,S_{r}}\mu
_{c|S_{1}|}\ldots \dfrac{\mu _{c|S_{r}|}}{d\beta }  \label{eq26}
\end{eqnarray}%
where in the first line we have used Eq. (\ref{eq22}) at $k=p+1$. The term $%
\mu _{p+1}\mu _{1}$ can be rewritten as 
\begin{equation}
\mu _{p+1}\mu _{1}=\sum\limits_{r=1}^{p+1}\sum\limits_{S_{1},\ldots
,S_{r}}\mu _{c|S_{1}|}\ldots \mu _{c|S_{r}|}\mu _{c|T|}  \label{eq27}
\end{equation}%
Here, to the set of natural numbers $\{1,\ldots ,p+1\}$ we added the number $%
p+2$ so that in the partition $S_{1},\ldots ,S_{r},T$ the $S$'s correspond
to all possible blocks partitioning the set $\{1,\ldots ,p+1\}$ and $T$ is
related to a fixed element, the number $p+2$. Evidently, $|T|=1$ and $\mu
_{c|T|}=\mu _{c1}=\mu _{1}$. Further, by induction conjecture one can use
Eq. (\ref{eq23}) for the cumulant derivatives in the second and next lines 
\begin{eqnarray}
\dfrac{\mu _{c|S_{1}|}}{d\beta } &=&-\mu _{c(k_{1}+1)}=-\mu _{c|S_{1}\cup
T|},  \notag \\
&\vdots &  \notag \\
\dfrac{\mu _{c}\left[ S_{r}\right] }{d\beta } &=&-\mu _{c(k_{r}+1)}=-\mu
_{c|S_{r}\cup T|},  \label{eq28}
\end{eqnarray}%
where $S_{1}\cup T,\ldots ,S_{r}\cup T$ denote that to a particular
partition of $\{1,\ldots ,p+1\}$ we add consecutively a fixed element $T=p+2$
to $S_{1},\ldots ,S_{r}$ blocks. Substituting Eqs. (\ref{eq27}) and (\ref%
{eq28}) into (\ref{eq26}), we get 
\begin{eqnarray}
\dfrac{d\mu _{c(p+1)}}{d\beta } &=&-\mu
_{p+2}+\sum\limits_{r=1}^{p+1}\sum\limits_{S_{1},\ldots ,S_{r}}\mu
_{c|S_{1}|}\ldots \mu _{c|S_{r}|}\mu _{c|T|}  \notag \\
&+&\sum\limits_{r=2}^{p+1}\sum\limits_{S_{1},\ldots ,S_{r}}\mu _{c|S_{1}\cup
T|}\ldots \mu _{c|S_{r}|}+\cdots  \notag \\
&+&\sum\limits_{r=2}^{p+1}\sum\limits_{S_{1},\ldots ,S_{r}}\mu
_{c|S_{1}|}\ldots \mu _{c|S_{r}\cup T|}  \label{eq29}
\end{eqnarray}%
Evidently, summing over $S_{1},\ldots ,S_{r}$ partitions of $\{1,\ldots
,p+1\}$ with inclusion of a fixed element $T$ is equivalent to 
\begin{equation*}
\sum\limits_{r=2}^{p+2}\sum\limits_{S_{1},\ldots ,S_{r}}
\end{equation*}%
where $S_{1},\ldots ,S_{r}$ is a partition of $\{1,\ldots ,p+2\}$ at $r\geq
2 $, so that Eq. (\ref{eq29}) can be rewritten as 
\begin{eqnarray}
\dfrac{d\mu _{c(p+1)}}{d\beta } &=&-\mu
_{p+2}+\sum\limits_{r=2}^{p+2}\sum\limits_{S_{1},\ldots ,S_{r}}\mu
_{c|S_{1}|}\ldots \mu _{c|S_{r}|}  \notag \\
&=&-\mu _{c(p+2)}  \label{eq30}
\end{eqnarray}%
Here, in transition to the second line, we employed Eq. (\ref{eq25}), where
a substitution $p\rightarrow p+1$ has been made. This completes the proof.

\section{Multivariate Cumulants and Proof of Theorem \protect\ref{T2}}

\label{App3}

Similar to Eq. (\ref{A0}), multivariate moments and cumulants are defined
via the moment generating function \cite{Ku:62} 
\begin{eqnarray}
M(t_{1},\ldots ,t_{p}) &=&\left\langle \exp \left(
\sum_{k=1}^{p}t_{k}H_{k}\right) \right\rangle =1+\sum_{k_{1},\ldots ,k_{p}}%
\dfrac{t_{1}^{k_{1}}\cdots t_{p}^{k_{p}}}{k_{1}!\cdots k_{p}!}\mu
_{k_{1},\ldots ,k_{p}}  \notag \\
&=&\exp \left[ \sum_{k_{1},\ldots ,k_{p}}\dfrac{t_{1}^{k_{1}}\cdots
t_{p}^{k_{p}}}{k_{1}!\cdots k_{p}!}\mu _{ck_{1},\ldots ,k_{p}}\right]
\label{C1}
\end{eqnarray}%
where $\mu _{k_{1},\ldots ,k_{p}}\equiv \left\langle H_{1}^{k_{1}}\cdots
H_{p}^{k_{p}}\right\rangle $ and summation runs over non-negative integers $%
k_{1},\ldots ,k_{p}$ except for $k_{1}=\ldots =k_{p}=0$. Using Eq. (\ref{C1}%
), one can derive explicit relations between $p$-dimensional moments and
cumulants, multivariate generalizations of Eqs. (\ref{A0}) and (\ref{A3});
see Appendix in \cite{Mee:57}. Following \cite{Mee:57}, it is convenient to
introduce compact notations for multi-indexes $\mathbf{k}=(k_{1},\ldots
,k_{p})$, as well as for the product of powers of multivariate variables $%
\mathbf{t}^{\mathbf{k}}=t_{1}^{k_{1}}\cdots t_{p}^{k_{p}}$, and product of
factorials $\mathbf{k}!=k_{1}!\cdots k_{p}!$. In these notations, Eq. (\ref%
{C1}) reads%
\begin{eqnarray}
\sum_{\mathbf{k}}\frac{\mathbf{t}^{\mathbf{k}}}{\mathbf{k}!}\mu _{\mathbf{k}%
} &=&\exp \left[ \sum_{\mathbf{k}}\frac{\mathbf{t}^{\mathbf{k}}}{\mathbf{k}!}%
\mu _{c\mathbf{k}}\right] -1  \notag \\
&=&\sum_{r=1}^{\infty }\frac{1}{r!}\sum_{\mathbf{n}_{1}\neq \mathbf{0}%
,\ldots ,\mathbf{n}_{r}\neq \mathbf{0}}\frac{\mathbf{t}^{\mathbf{n}%
_{1}+\ldots +\mathbf{n}_{r}}}{\mathbf{n}_{1}!\cdots \mathbf{n}_{r}!}\mu _{c%
\mathbf{n}_{1}}\cdots \mu _{c\mathbf{n}_{r}}  \label{C2}
\end{eqnarray}%
In the second line, the exponential function has been expanded in Taylor's
series and multi-indexes in the inner-sum are non-negative such that $%
\mathbf{n}_{1}\neq \mathbf{0},\ldots ,\mathbf{n}_{r}\neq \mathbf{0}$.
Differentiation of Eq. (\ref{C2}) by applying the multi-index differential
operator%
\begin{equation}
D_{\mathbf{t}}^{\mathbf{k}}=\frac{\partial ^{|\mathbf{k}|}}{\partial
t_{1}^{k_{1}}\cdots \partial t_{p}^{k_{p}}}  \label{C3}
\end{equation}%
to the both sides and then putting $\mathbf{t}=\mathbf{0}$ yields a
multivariate analogue of (\ref{A3})%
\begin{equation}
\mu _{\mathbf{k}}=\sum_{r=1}^{|\mathbf{k}|}\frac{1}{r!}\sum_{\substack{ 
\mathbf{n}_{1}\neq \mathbf{0},\ldots ,\mathbf{n}_{r}\neq \mathbf{0}  \\ 
\mathbf{n}_{1}+\ldots +\mathbf{n}_{r}=\mathbf{k}}}\frac{\mathbf{k}!}{\mathbf{%
n}_{1}!\cdots \mathbf{n}_{r}!}\mu _{c\mathbf{n}_{1}}\cdots \mu _{c\mathbf{n}%
_{r}}  \label{C4}
\end{equation}%
where $|\mathbf{k}|=k_{1}+\ldots +k_{p}$.

In multi-dimensional case, Eq. (\ref{A4a}) remains valid if the single index 
$k$ is replaced by a multi-dimensional one $\mathbf{k}$, and partitioning $%
\pi _{r}$ of the set $\{1,2,\ldots ,k\}$ is replaced by partitioning of the
ordered set of multi-indexes $(\{1,\ldots ,k_{1}\};\{1,\ldots
,k_{2}\};\ldots ;\{1,\ldots ,k_{p}\})$: 
\begin{eqnarray}
\mu _{\mathbf{k}} &=&\sum_{r=1}^{|\mathbf{k}|}\sum_{S_{1},\ldots ,S_{r}\in
\pi _{r}}\mu _{c|S_{1}|}\cdots \mu _{c|S_{r}|}  \notag \\
&=&\sum_{r=1}^{|\mathbf{k}|}\sum_{\substack{ \mathbf{n}_{1}\leq \cdots \leq 
\mathbf{n}_{r}  \\ \mathbf{n}_{1}+\cdots +\mathbf{n}_{r}=\mathbf{k}}}%
\sum_{|S_{1}|=\mathbf{n}_{1},\ldots ,|S_{r}|=\mathbf{n}_{r}}\mu _{c\mathbf{n}%
_{1}}\cdots \mu _{c\mathbf{n}_{r}}  \label{C5}
\end{eqnarray}%
where $S_{1},\ldots ,S_{r}$ are all possible blocks into which the set of
multi-indexes can be divided. The summation over blocks in the second line
is organized as follows. The first summation over $r$ defines the number of
blocks $S_{1},\ldots ,S_{r}$ partitioning the set of multi-indexes. The
second summation runs over all possible sizes of blocks such that $\mathbf{n}%
_{1}+\cdots +\mathbf{n}_{r}=\mathbf{k}$. In components, if $\mathbf{n}%
_{i}=(n_{i1},n_{i2},\ldots ,n_{ip})$, $i=1,\ldots ,r$ [the first subindex
labels the block's number, the second one is due to variable's number], we
have constraining equations 
\begin{equation}
\mathbf{n}_{1}+\cdots +\mathbf{n}_{r}=\mathbf{k}\quad \Leftrightarrow \quad %
\begin{cases} n_{11}+\cdots +n_{r1}=k_1 \\ n_{12}+\cdots +n_{r2}=k_2 \\
\hdotsfor{1} \\ n_{1p}+\cdots +n_{rp}=k_p \end{cases}
\end{equation}%
and inequalities specifying the partial ordering conditions 
\begin{equation}
\mathbf{n}_{1}\leq \cdots \leq \mathbf{n}_{r}\quad \Leftrightarrow \quad %
\begin{cases} n_{11}\le n_{21}\le\cdots\le n_{r1} \\ n_{12}\le
n_{22}\le\cdots\le n_{r2} \\ \hdotsfor{1} \\ n_{1p}\le n_{2p}\le\cdots\le
n_{rp} \end{cases}
\end{equation}%
The inner summation over all possible blocks at fixed sizes results in 
\begin{equation}
\sum_{|S_{1}|=\mathbf{n}_{1},\ldots ,|S_{r}|=\mathbf{n}_{r}}\mu _{c\mathbf{n}%
_{1}}\cdots \mu _{c\mathbf{n}_{r}}=\dfrac{\mathbf{k}!}{\mathbf{n}_{1}!\cdots 
\mathbf{n}_{r}!}\mu _{c\mathbf{n}_{1}}\cdots \mu _{c\mathbf{n}_{r}}
\label{C8}
\end{equation}%
where 
\begin{equation}
\dfrac{\mathbf{k}!}{\mathbf{n}_{1}!\cdots \mathbf{n}_{r}!}=\dfrac{%
k_{1}!\cdots k_{p}!}{n_{11}!\cdots n_{1p}!\cdots n_{r1}!\cdots n_{rp}!}
\label{C9}
\end{equation}%
Observe that the multi-index-multinomial coefficient (\ref{C9}) in the
right-hand side of (\ref{C8}) is the same as in (\ref{C4}). As in Appendix %
\ref{App1}, due to symmetry of the summed functions with respect to
permutations of multi-indexes $\mathbf{n}_{1},\ldots ,\mathbf{n}_{r}$, we
obtain that Eqs. (\ref{C4}) and (\ref{C5}) are equivalent representations
for multivariate moments. The latter representation is helpful in proving
the multivariate analogue of the theorem on cumulant's derivative.

The proof of Theorem \ref{T2} is by induction and quite similar to the proof
of Theorem \ref{T1} done in previous Appendix. The only complication is due
to multi-index nomenclature.

\newpage

\end{document}